\documentclass[12pt,preprint]{emulateapj}
\journalinfo{Accepted for publication in The Astrophysical Journal (July 12, 2011)}
\shorttitle{VLA Survey of the CDFS. V. Evolution and Luminosity Functions}
\shortauthors{Padovani et al.}

\begin{document}
\def\proptoapprox{\mathrel{\hbox{\rlap{\hbox{\lower4pt\hbox{$\sim$}}}\hbox{$\propto$}}}}
\newdimen\digitwidth
\setbox0=\hbox{2}
\digitwidth=\wd0
\catcode `#=\active
\def#{\kern\digitwidth}

\title{THE VLA SURVEY OF THE CHANDRA DEEP FIELD SOUTH. 
V. EVOLUTION AND LUMINOSITY FUNCTIONS OF SUB-MJY RADIO SOURCES AND 
THE ISSUE OF RADIO EMISSION IN RADIO-QUIET AGN}

\author{P. Padovani}
\affil{European Southern Observatory, Karl-Schwarzschild-Str. 2,
D-85748 Garching bei M\"unchen, Germany}
\email{ppadovan@eso.org}

\author{N. Miller} 
\affil{Department of Astronomy, University of Maryland,
  College Park, MD 20742-2421, USA}

\author{K. I. Kellermann}
\affil{National Radio Astronomy Observatory, 520 Edgemont Road, 
 Charlottesville, VA 22903-2475, USA}

\author{V. Mainieri, P. Rosati}
\affil{European Southern Observatory, Karl-Schwarzschild-Str. 2,
D-85748 Garching bei M\"unchen, Germany}

\author{P. Tozzi}
\affil{INAF, Osservatorio Astronomico di Trieste, Via G. B. Tiepolo 
11, I-34131, Trieste, Italy}

\begin{abstract}
  We present the evolutionary properties and luminosity functions of the
  radio sources belonging to the Chandra Deep Field South VLA survey, which
  reaches a flux density limit at 1.4 GHz of $43~\mu$Jy at the field center
  and redshift $\sim 5$, and which includes the first radio-selected
  complete sample of radio-quiet active galactic nuclei (AGN). We use a
  new, comprehensive classification scheme based on radio, far- and
  near-IR, optical, and X-ray data to disentangle star-forming galaxies
  from AGN and radio-quiet from radio-loud AGN. We confirm our previous
  result that star-forming galaxies become dominant only below 0.1 mJy. The
  sub-millijansky radio sky turns out to be a complex mix of star-forming galaxies
  and radio-quiet AGN evolving at a similar, strong rate; non-evolving
  low-luminosity radio galaxies; and declining radio powerful ($P \ga 3
  \times 10^{24}$ W~Hz$^{-1}$) AGN. Our results suggest that radio emission
  from radio-quiet AGN is closely related to star formation. The detection
  of compact, high brightness temperature cores in several nearby
  radio-quiet AGN can be explained by the co-existence of two components,
  one non-evolving and AGN-related and one evolving and
  star-formation-related. Radio-quiet AGN are an important
  class of sub-millijansky sources, accounting for $\sim 30\%$ of the sample and
  $\sim 60\%$ of all AGN, and outnumbering radio-loud AGN at $\la 0.1$
  mJy. This implies that future, large area sub-millijansky surveys, given the 
  appropriate ancillary multi-wavelength data, have the potential of  
  being able to assemble vast samples of radio-quiet AGN by-passing the problems of
  obscuration, which plague the optical and soft X-ray bands.
  \end{abstract}


%
%________________________________________________________________

\section{Introduction}\label{intro}

The relationship between star formation and AGN in the Universe is one of
the hottest topics of current extragalactic research, at two different
levels. On cosmological scales, the growth of supermassive black holes in
AGN appears to be correlated with the growth of stellar mass in galaxies
\citep[e.g.,][]{mer08}. On nuclear scales, the accreting gas feeding the
black hole at the center of the AGN might trigger a starburst. The black
hole, through winds and jets, can in turn feed energy back to its
surroundings, which can compress the gas and therefore accelerate star
formation but can also blow it away, thereby stopping accretion and star
formation altogether. 
The general consensus is that nuclear activity plays a major role in the
co-evolution of supermassive black holes and galaxies through the so-called
``AGN Feedback'' and indeed radio emission from AGN has been recently
suggested to play an important role in galaxy evolution
\citep{cro06}. Moreover, radio observations afford a view of the Universe
unaffected by the absorption, which plagues observations made at most other
wavelengths, and therefore provide a vital contribution to our
understanding of this co-evolution. These two points imply that studies of
the evolution of star-forming galaxies (SFG) and AGN in the radio band
should provide a better understanding of the link between the two
phenomena. These are obviously done best by reaching relatively faint ($\la
1$ mJy) flux densities, and hence the importance of characterizing the
radio faint source population.

After years of intense debate, the contribution to the sub-millijansky population from
synchrotron emission resulting from relativistic plasma ejected from
supernovae associated with massive star formation in galaxies appears not
to be overwhelming, at least down to $\sim 50~\mu$Jy, contrary to the
(until recently) most accepted paradigm. Our deep ($S_{\rm 1.4GHz} \ge
43~\mu$Jy) radio observations with the NRAO Very Large Array (VLA) of the
Chandra Deep Field South (CDFS), complemented by a variety of data at other
frequencies, imply in fact a roughly 50/50 split between SFG and AGN
\citep{pad09}, in broad agreement with other recent papers
\citep[e.g.,][]{sey08,smo08}. 

The purpose of this paper is to study the evolution and luminosity
functions (LFs) of sub-millijansky radio sources through the VLA-CDFS sample. Apart
from the topics mentioned above, this is important also for other issues,
including: 

\begin{enumerate}

\item predictions for the source population at radio flux densities $<
  1~\mu$Jy, which are relevant, for example, for the Square Kilometre Array
  (SKA). All existing estimates, in fact, had to rely, for obvious reasons,
  on extrapolations and are based on high flux density samples. This
  affects particularly the highest redshifts, which 
  can better be probed at fainter flux densities;

\item the radio evolution of radio-quiet AGN. No radio-selected sample of
  radio-quiet AGN is currently available and this is badly needed to shed
  light on the mechanism behind their radio emission and allow a proper
  comparison with radio-loud quasars;

\item the fact that number counts by themselves do not necessarily reflect
  the relative intrinsic abundance of astrophysical sources, which requires
  the determination of the evolution and LF \citep{pad07}.

\end{enumerate}

We note that the evolution and LFs of sub-millijansky radio sources have been
studied, so far, only in two fields: the COSMOS field \citep{smo09a,smo09b}
and the Deep {\it Spitzer} Wide-area InfraRed Extragalactic (SWIRE) field
\citep{str10}, in both cases up to a maximum redshift of $1.3$ and without
differentiating between radio-quiet and radio-loud AGN. Source
classification was based on a rest-frame optical color scheme and on
spectral energy distribution (SED) fitting to photometric data covering the
UV to near-IR range respectively.

We define as AGN sources in which most of the energy is produced through
physical processes other than the nuclear fusion that powers stars. In
practice, this means that electromagnetic emission most likely related,
directly or indirectly, to a supermassive black hole is predominant in at
least one band. A small fraction of AGN have, for the same optical power,
radio powers three to four orders of magnitude higher than the rest. These are
called ``radio-loud'' quasars and most of the energy they emit is
non-thermal and is associated with powerful relativistic jets, although
thermal components associated with an accretion disk may also be observed,
especially in the optical/UV band. Radio galaxies are also characterized by
strong radio jets (manifested also through radio lobes), typically laying
in or near the plane of the sky, and a fraction of them (the most powerful
ones) are thought to be radio-loud quasars, which have instead their jets
oriented with a small angle to the line of sight \citep[e.g.,][]{up95}.
We define as
``radio-quiet'' AGN in which jets are either not present or make a tiny
contribution to the total energy budget over the whole electromagnetic
spectrum, which is dominated by thermal emission. All other AGN we call
radio-loud. Note that radio-quiet AGN are not radio-silent. Indeed, the
radio power of many low-luminosity radio galaxies, the so-called
Fanaroff-Riley (FR) Is (``low-power radio-loud AGN'' according to our
nomenclature) overlap with that of radio-quiet AGN, which can generate some
confusion and requires great care during the classification
process, which needs to involve also the X-ray and far-IR bands (Section~\ref{revised}). 
However, the two classes are physically distinct (see
Sections~\ref{2_lowrl} and \ref{astro}), although the origin of radio
emission in radio-quiet AGN is still not clear (but see
Section~\ref{rq_emission}).

Translating these high-level definitions into a classification scheme
requires a wealth of multi-wavelength data, which were described in our
previous papers. \cite{kel08} (Paper I) presented the radio data of
the VLA-CDFS sample, together with optical images and X-ray counterparts,
while \cite{mai08} (Paper II), discussed the optical and near IR
counterparts to the observed radio sources and, based on rest-frame colors
and the morphology of the host galaxies, found evidences for a change in
the sub-millijansky radio source population below $\approx 80$
microJy. \cite{toz09} (Paper III), dealt with the X-ray properties, while
\cite{pad09} (Paper IV), discussed the source population.  This turned out
to be made up of SFG and AGN at roughly equal levels, with the AGN
including radio galaxies, mostly low-power (FR Is), and a significant
($\sim 50\%$) radio-quiet component. Paper IV made also clear that the
``standard'' definitions of radio-loudness, based on radio-to-optical flux
density ratios, $R$, and radio powers, were insufficient to identify
radio-quiet AGN when dealing with a sample, which included also
star-forming and radio-galaxies, as both classes are or can be
(respectively) characterized by low $R$ and radio powers as well. 
$R$, for example, is useful for quasar samples, where it can be assumed
that the optical flux is related to the accretion disk, but obviously loses its
meaning as an indicator of jet strength if both the radio and the optical
band are dominated by jet emission, as might be the case in FR Is
\citep{chi99}.  Source classification in Paper IV was then based on radio,
optical, and X-ray data, and was meant to provide a robust upper limit to
the fraction of SFG at sub-millijansky levels. SFG candidates were
selected based on their (low values of) $R$, (low) radio power,
(non-elliptical or S0) optical morphology, and (low) X-ray power ($L_{\rm
  x}$). The fact that X-ray upper limits above the AGN threshold ($10^{42}$
ergs s$^{-1}$) were also included was conservative in the sense that it
maximized the number of SFG, as some of these sources could still be
AGN. Furthermore, the selection of radio-quiet AGN candidates was only
approximate, as it was based solely on $R$ and $L_{\rm x}$ and suffered
from uncertainties in the optical K-correction, possible contamination by
radio galaxies, and the exclusion of X-ray upper limits (see Paper IV for
details). In order to deal with the evolution and luminosity functions of
the various classes of sources, we need to refine our classification. In
particular, the CDFS field has been observed by {\it Spitzer} and therefore
near- (Section \ref{nearIR}) and mid/far-IR (Section \ref{farIR}) data are
available for our sample.  The source classification used in this paper
relies then on a combination of radio, IR, optical, and X-ray data
(Section~\ref{revised}).

%About half of these AGN are radio-quiet, that is of the type normally found in 
%optically selected samples and characterized by relatively low radio-to-optical 
%flux density ratios and radio powers, as compared to radio quasars. These objects 
%represent a small minority above 1 mJy.

Section~2 describes the updated classification of the VLA-CDFS sample, 
while Section~3 studies its evolution. Section~4 derives the LFs for 
various classes, while Section~5 discusses our results. Finally, 
Section~6 summarizes our
conclusions. Throughout this paper spectral indices are defined by $S_{\nu}
\propto \nu^{-\alpha}$ and the values $H_0 = 70$ km s$^{-1}$ Mpc$^{-1}$,
$\Omega_{\rm M} = 0.3$, and $\Omega_{\rm \Lambda} = 0.7$ have been
used. 

\section{The updated sample classification}\label{sample}

\subsection{Redshifts}\label{newz}

Our sample includes all VLA-CDFS sources with reliable optical
counterparts and eight empty fields, for total of 256 objects, 193 of which
belong to a complete sample\footnote{Four more sources belonging to the
complete sample have very uncertain counterparts (see Paper II) and for
one other source, very close to a bright star, we could not get reliable
photometry. The inclusion of these sources in any of the classes
described below would change our results by much less than $1\sigma$.} (see
Paper IV for details). 92\% of the sources in the complete sample
now have redshift information (74\% spectroscopic) 
as compared to only 77\% 
in Paper IV. We have included in this paper spectroscopic redshifts from a
dedicated follow-up program performed with the VIMOS spectrograph at the
VLT (Bonzini et al., in preparation).  We also used recently published
redshifts for the counterparts of Chandra sources in this field
\citep{tre09,si10} and the photometric redshifts published by
the Multiwavelength Survey by Yale-Chile (MUSYC) \citep{car10}, which are
based on 32 photometric bands.

\begin{figure}%1
%\plotone{zvmag.eps}
\centerline{\includegraphics[width=9.5cm]{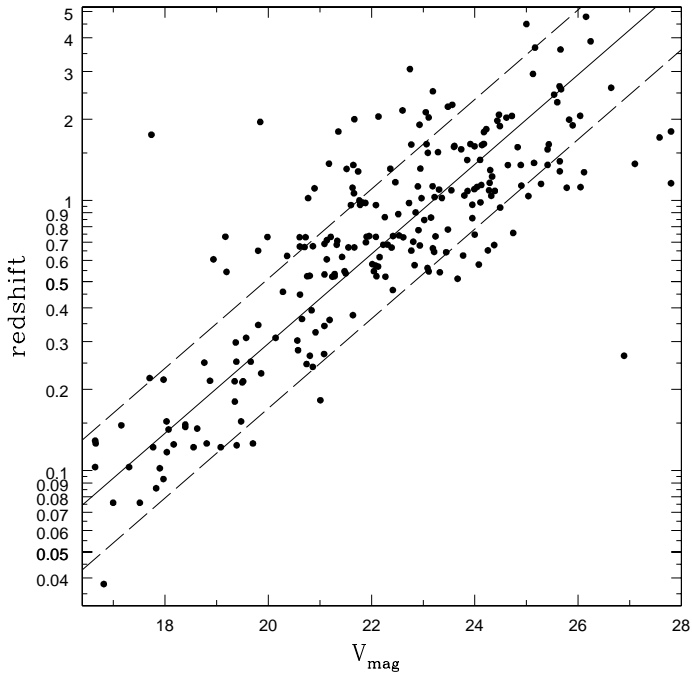}}
%\plotone{fig1.eps}
\caption{Redshift versus $V_{\rm mag}$ for our sources with redshift
  information. The solid line is the best fit, while the dashed lines
  represent the scatter. See text for details.}
\label{zvmag}
\end{figure}

As shown in Fig. \ref{zvmag}, redshift is strongly correlated with
magnitude, albeit with some scatter. The best and simplest approach to
estimate the redshift for the 16 objects in the complete sample without
observed redshifts is then to derive it from their magnitude by using the
relationship shown in the figure (solid line), that is $\log z = 0.166
V_{\rm mag} - 3.85$. This was derived applying to the whole sample the
ordinary least-square bisector method \citep{iso90}, which treats the
variables symmetrically. Including only spectroscopic redshifts, or only
the complete sample, or excluding sources with likely AGN contamination in
the optical band \citep[based on][]{szo04}, all give relations within
$1\sigma$ from the adopted one. The effect of this assumption on our
results is discussed in Section~\ref{missingz}.

Note that while objects well to the left of the correlation can be
explained as having an AGN component in the optical band, the single source
in the lower right part of the diagram is $\approx 7$ magnitudes fainter
than the average and therefore well into the dwarf galaxy regime. However,
its photometry is affected by its closeness to a bright star, which might
explain at least in part its faintness.

%As such ... CHECK WITH NORRIS  \cite{nor06} 

%study effect of estimated z (assume z = $\langle z \rangle$ and then larger)

\subsection{Near-IR data}\label{nearIR}

The usage of {\it Spitzer} Infrared Array Camera \citep[IRAC;][]{faz04}
colors to identify AGN has been discussed at length in the literature
\citep[e.g.,][]{la04,hat05,ste05,saj05,car08}. Although it is by now
evident that only some classes of extragalactic sources occupy restricted
regions of parameter space in such plots, it is nevertheless also clear
that there are broad trends which can be used to, for example, identify
possible misclassifications.

\begin{figure}%2
%\plotone{IRAC.eps}
\centerline{\includegraphics[width=9.5cm]{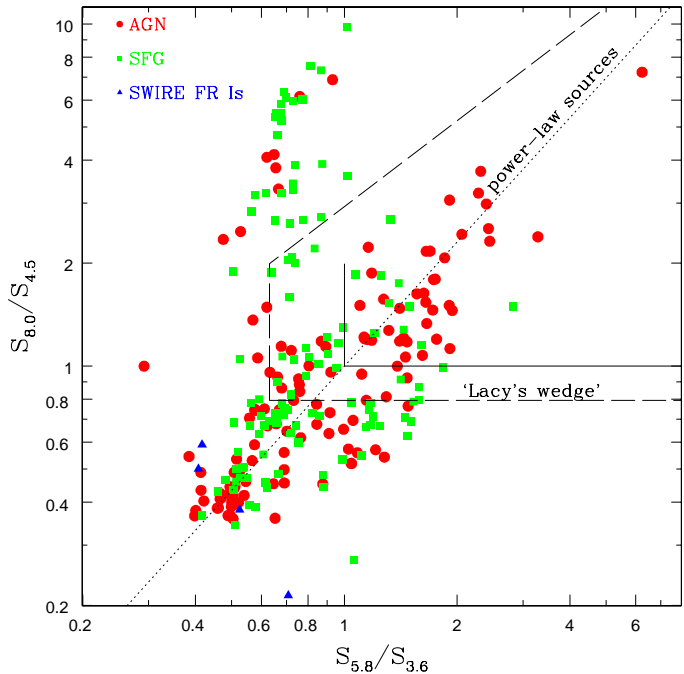}}
%\plotone{fig2.eps}
\caption{IRAC color-color plot for the AGN and SFG candidates selected in Paper IV 
  and four FR Is from the
  SWIRE field. The dotted line indicates the locus of sources whose
  spectrum can be described as a power-law over the four IRAC bands. The
  dashed lines indicate the so-called ``Lacy's wedge'', which is where most
  AGN are thought to lie. The solid lines denote a more restrictive region,
  which takes into account the fact that for $z > 0.5$ PAH- and
  star-light-dominated sources can be inside ``Lacy's wedge''
  \citep{das09}. See text for details.}
\label{IRAC}
\end{figure}

Fig. \ref{IRAC} plots the IRAC flux density ratios $S_{8.0}/S_{4.5}$
vs. $S_{5.8}/S_{3.6}$ for our sources classified as in Paper IV, where the
flux densities refer to all four IRAC channels at 3.6, 4.5, 5.8, and 8.0
$\mu$m. The IRAC data come from the {\it Spitzer} IRAC/MUSYC Public Legacy
survey in the Extended-CDFS \citep[SIMPLE;][]{dam11}. We cross-correlated
the SIMPLE catalogue with the VLA-CDFS sources, accepting matches
with separations less than 2$^{{\prime}{\prime}}$. The SIMPLE catalogue has
convolved the images associated with each IRAC channel to match that of
channel 4 (8.0 $\mu$m), the one with the lowest resolution, so that
reasonably accurate colors may be obtained from the four IRAC bands. We
have used total fluxes and applied the prescribed normalization to produce 
flux densities in $\mu$Jy. Fig. \ref{IRAC} shows the following:
(a) most AGN candidates fall around the locus of sources whose mid-IR spectrum 
can be characterized by a single power law (dotted line);
a significant number of AGN is also within the so-called ``Lacy's wedge'' 
(dashed lines), which
is where most unobscured, broad-lined (type 1) AGN are thought to lie
\citep{la04}. Note that highly obscured sources might also occupy that
region of parameter space \citep[e.g.,][]{das09,pra10}; (b) most SFG candidates 
are distributed in a vertical band centered around $S_{5.8}/S_{3.6} \sim 0.6 -
0.8$, which is where polycyclic aromatic hydrocarbon (PAH)- and
starlight-dominated sources are expected to lie \citep[e.g.,][]{saj05}; (c)
we also plot four ``bona fide'' FR I from the SWIRE field \citep{var08},
which fall in the region where galaxies with an old stellar population are 
located \citep[e.g.,][]{saj05}. We have quite a few sources
in the same area, which is consistent with one of the main results of Paper
IV, that is the dominance of low-luminosity radio galaxies amongst radio-loud
AGN.

This is reassuring and shows that the SFG/AGN division derived in Paper IV
is overall correct. The most interesting features in Fig. \ref{IRAC},
however, are the exceptions to the above, namely: (a) the eight AGN
candidates in the top left part of the diagram; these are all but one at
low redshift ($z \le 0.25$), low radio power ($\log P_{1.4GHz} \le 22.6$),
low X-ray ($2 - 10$ keV) power ($\log L_{\rm x} \le 41.6$) sources, which
had been classified as AGN solely because their optical morphology was S0
(5) or elliptical (2). A closer look at their images shows that two of them
show (weak) signs of spiral arms and four more (all S0) have only
low-resolution Wide Field Imager (WFI) data, which means that the presence
of spiral arms cannot be excluded. Their location in the PAH-dominated
region \citep{saj05} suggests a re-classification as SFG for all of them
apart from one AGN with two spiral galaxies at a distance of $\sim
3^{{\prime}{\prime}}$, which means its IRAC flux is most likely
contaminated (its rest-frame radio-to-optical flux density ratio is also
$\sim 2$, which is typical of radio-loud AGN: see below); (b) the ten SFG
candidates with $S_{5.8}/S_{3.6} > 1$ and $1 < S_{8.0}/S_{4.5} < 3$; this
is more restrictive than the Lacy's wedge as it takes into account the fact
that for $z > 0.5$ PAH- and star-light-dominated sources can be inside that
wedge \citep{das09}. Most of these sources have X-ray upper limits larger
than $10^{42}$ erg/s, which makes sense since this was one of the reasons
they were classified as SFG in the first place.
The location of these sources
suggests a re-classification as AGN. In summary, seven sources were re-classified
from AGN to SFG and ten sources previously classified SFG are now classified as AGN.

\subsection{Far-IR data}\label{farIR}

It is well know that the global far-IR and radio emission are tightly and
linearly correlated in star-forming systems \citep[e.g.,][and references
  therein]{sar10}. This is usually expressed through the so-called $q$
parameter, that is the logarithm of the ratio of far-IR to radio flux
density, as defined by \cite{hel85}. We take advantage of the
relatively narrow dispersion of $q$ for star-forming systems to further
refine our SFG/AGN separation and also to improve on our radio-quiet --
radio-loud AGN division, as the latter do not follow the IR - radio
correlation typical of SFG \citep[e.g.,][]{so91,sar10}. This is vital 
to separate radio-quiet AGN from radio galaxies, as $R$ is not very useful 
in this case (Section~\ref{intro}) and, like radio-quiet AGN, radio galaxies can also have 
relatively large X-ray powers.

We have used a catalog of $70~\mu$m Multiband Imaging Photometer for {\it
  Spitzer} (MIPS) flux densities from the Far- Infrared Deep Extragalactic
Survey (FIDEL; Dickinson et al. in preparation) for our evaluation of
$q$. We cross correlated the VLA-CDFS radio sources with the FIDEL
catalogue using a radius of 8$^{{\prime}{\prime}}$ (about half the {\it
  Spitzer} $70~\mu$m point-spread-function). For those sources undetected
by the FIDEL survey (but still within the FIDEL coverage), we assume an
upper limit of 2.5 mJy as this is approximately the 5$\sigma$ survey
limit. To these data we add $24~\mu$m flux densities from the Great
Observatories Origins Deep Survey (GOODS) whenever available, and thus we
obtain SEDs sampled at up to eight wavelengths: 20 and 6 cm in the radio
from our VLA surveys; $70~\mu$m and $24~\mu$m in the IR from FIDEL and
GOODS; and $8.0~\mu$m, $5.6~\mu$m, $4.5~\mu$m, and $3.6~\mu$m in the
near-IR from SIMPLE.

We then proceeded to find the template SED from the \cite{dal01} SFG models
that best matches the {\it Spitzer} data. We use the source redshifts to
place each of the 64 models into the observed frame for that source, and
set the normalization by requiring that each model SED pass through the
measured $70~\mu$m flux density for that galaxy. This, in effect, places an
extra weight on $70~\mu$m data since it is our only measurement of the
smooth modified blackbody portion of the SED. We then select the model
which minimizes the least squares fit to the photometry of the four IRAC
channels and the MIPS $24~\mu$m data (when available). Once the
best-fitting model has been selected, we derive the rest-frame $60~\mu$m
and $100~\mu$m flux densities to determine FIR, the total far-IR flux
between $42.5~\mu$m and $122.5~\mu$m \citep{hel85}

\begin{equation}
FIR = 1.26 \times 10^{-14}~[2.58f_{60\mu m} + f_{100\mu m}] ~{\rm W~m}^{-2}
\label{eq_FIR}
\end{equation}
where the flux densities, $f$, are in Jy. Similarly, we convert the
observed 1.4~GHz radio emission to the rest frame using its
measured spectral index between 1.4~GHz and 4.86~GHz, where available ($\sim 
80\%$ of the sample: see Paper I), or
assuming a spectral index $\alpha_{\rm r} = 0.7$ (the mean of the sample) otherwise. 
The value of
$q$ is then calculated as the logarithm of the ratio of far-IR to 1.4~GHz
flux density:

\begin{equation}
q = log{[(FIR/3.75 \times 10^{12})/S_{1.4GHz}]}
\label{eq_q}
\end{equation}
where the numeric factor is the frequency in Hz corresponding to a
wavelength of $80~\mu$m.

Given the large fraction ($\sim 50\%$) of upper limits on $q$, one cannot
readily look for a bimodality in its distribution to define a dividing line
between star-forming and non star-forming sources. The median of the
detections should be however quite well defined, as its value is 2.16 and
most upper limits are below 2.2. Since 96\% of the detections above the
median are below 2.64 and assuming a symmetric distribution, one finds a
lower end at around $2.16-(2.64-2.16) \sim 1.7$. We then assume in the
following that sources characterized by $q \ge 1.7$ are
star-formers\footnote{Our results are only weakly dependent on this
  choice. For example, if we defined as star-formers sources with $q \ge
  1.8$ our SFG complete sample would only lose three objects, a 4\% effect
  (see Tab. \ref{tabveva}).} (upper limits above this value excluded). This is
the same dividing value assumed by \cite{mac99}.
Twenty-two of our candidate SFG have $q < 1.7$ and therefore cannot be
star-forming systems. These were then re-classified as radio-loud
AGN. These sources fall in the region where passive galaxies are found
in the IRAC color-color plot, which is consistent with this
re-classification, given that most of our radio-loud AGN should be radio
galaxies.

Finally, eight radio-quiet AGN candidates were found to have $q < 1.7$, while nineteen
radio-loud ones had $q \ge 1.7$, which reflects the approximation of our
previous classification. These objects were re-classified as radio-loud and
radio-quiet respectively. 

\subsection{Revised classification}\label{revised}

\begin{figure}%3
\epsscale{1.35} 
\plotone{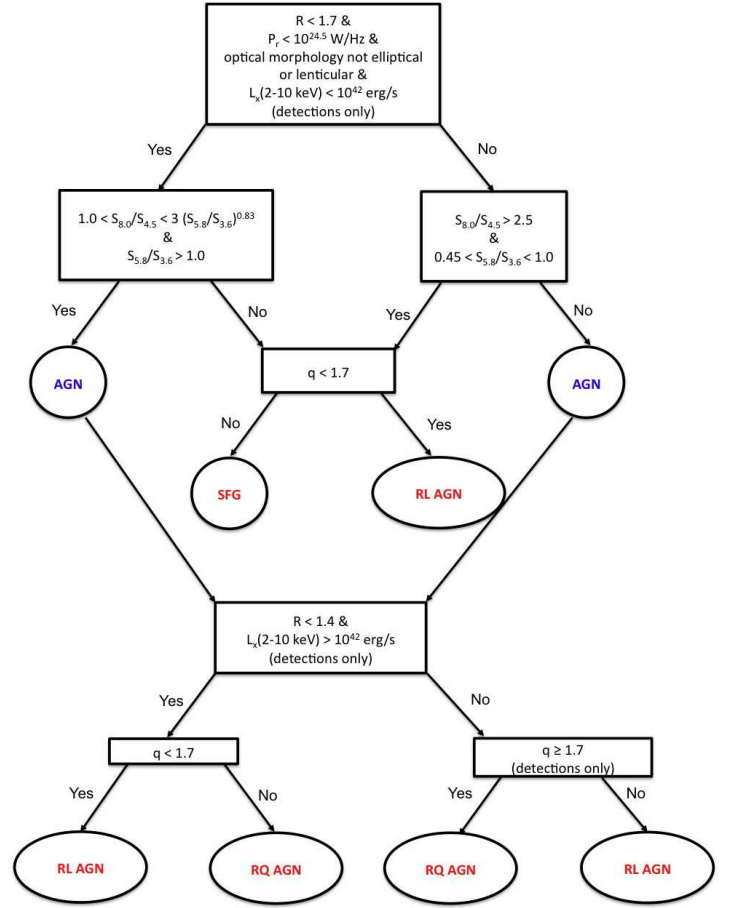}
%\plotone{flowchart_sfg_agn.new.eps}
\caption{A flow chart of our classification scheme. See text for details.}
\label{flow}
\end{figure}

To summarize, based on the results presented in Paper IV and in the 
previous sub-sections, our candidate star-forming galaxies are defined as
fulfilling the following initial requirements:

\begin{enumerate}
\item $R = \log (S_{\rm 1.4GHz}/S_{\rm V}) < 1.7$ (where $S_{\rm V}$ is the
V-band flux density)

\item $P_{\rm r} < 10^{24.5}$ W Hz$^{-1}$

\item optical morphology different from elliptical or lenticular 

\item $L_{\rm x} (2-10~\rm keV) < 10^{42}$ ergs s$^{-1}$ for X-ray
  detections, no limit otherwise.
\end{enumerate}

As discussed in Paper IV, the first two criteria include $\sim 90\%$ of
spirals and irregulars, the third one excludes sources not associated with
star formation at our redshifts ($\langle z \rangle \sim 1.1$), while the fourth one
excludes AGN. These are then supplemented by the following additional
requirements, which can overrule the previous ones if necessary:

\begin{enumerate}
\setcounter{enumi}{4}

\item IRAC constraints: the region of parameter space defined by
  $S_{5.8}/S_{3.6} > 1$ and $1 < S_{8.0}/S_{4.5} \la 3~(S_{5.8}/S_{3.6})^{0.83}$ 
  (AGN region) is
  excluded; sources not classified as SFG by the previous criteria but with
  $0.45 < S_{5.8}/S_{3.6} < 1.0$ and $S_{8.0}/S_{4.5} > 2.5$ (PAH-dominated
  region) are also included (Section~\ref{nearIR})

\item MIPS constraints: $q \ge 1.7$, upper limits above this value excluded
  (Section~\ref{farIR})

\end{enumerate}

Note that constraints number 2 and 3 have become almost irrelevant for our
classification given these two new requirements. Objects not fulfilling
this sequence of criteria are considered to be AGN. Radio-quiet AGN are
defined initially as follows:

\begin{enumerate}
\item $R < 1.4$ 

\item $L_{\rm x} (2-10~\rm keV) > 10^{42}$ ergs s$^{-1}$ (detections only)
  
\end{enumerate}

As discussed in Paper IV, the first criterion is the ``classical''
definition of radio-quiet AGN converted to the 1.4 GHz and V bands. These
are then supplemented by the following additional requirements, which can
overrule the previous ones if necessary:

\begin{enumerate}
\setcounter{enumi}{2}

\item IRAC constraints: the region of parameter space defined by 
$S_{8.0}/S_{4.5} > 2.5$ and $0.45 < S_{5.8}/S_{3.6} < 1.0$ (PAH-dominated 
region) is excluded (Section~\ref{nearIR})

\item MIPS constraints: $q \ge 1.7$, upper limits above this value excluded
  (Section~\ref{farIR})

\end{enumerate}

AGN not fulfilling this sequence of criteria are considered to be
radio-loud. Fig. \ref{flow} summarizes our classification scheme.  We note
that, while it is relatively simple to distinguish SFG from radio-loud AGN
due to their different $q$ values, the situation is more complicated when
one has to differentiate SFG from radio-quiet AGN. This is done also based
on the location on the IRAC color-color plot, which highlights the obvious
outliers, but mostly on the basis of X-ray power. Since many SFG still have
upper limits on $L_{\rm x} (2-10~\rm keV) > 10^{42}$ ergs s$^{-1}$, we
cannot exclude that more radio-quiet AGN are present in our sample, 
especially amongst the SFG with the highest limits on $L_{\rm x} (2-10~\rm
keV)$, which tend to be at higher redshifts. The inclusion of deeper X-ray
data in our analysis will help us sort out this issue.

It is instructive to see how representative local sources get classified by
our scheme. To this aim, we have used the NASA/IPAC Extragalactic Database
(NED) and NASA's Astrophysics Data System (ADS) to get multi-wavelength
data for a few objects. For example, NGC 1068, the prototype Seyfert 2
galaxy, residing in a spiral host, has low $R$, $P_{\rm r}$, and $L_{\rm x}
(2-10~\rm keV)$ values. Coupled with a $q > 1.7$ it would then be
classified as an SFG but its location on the IRAC color-color plot puts it
firmly with the radio-quiet AGN. NGC 1052, an elliptical galaxy often
classified as a low-ionization nuclear emission-line region (LINER), has
also low $R$, $P_{\rm r}$, and $L_{\rm x} (2-10~\rm keV)$ values but its
low $q$ makes it a radio-loud AGN. M 82, the prototype starburst galaxy,
again has low $R$, $P_{\rm r}$, and $L_{\rm x} (2-10~\rm keV)$ values but
its $q > 1.7$ and location on the IRAC diagram classify it as a SFG. And
NGC 1275, a cD (radio) galaxy at the centre of the Perseus cluster, which
in the literature has been classified, amongst other things, as a Seyfert
1.5 and a blazar, with its high $R$, $P_{\rm r}$, and very low $q$ is
undoubtedly a radio-loud AGN.

\begin{deluxetable*}{lcccccccc}
\tabletypesize{\scriptsize}
\tablecaption{Euclidean normalized 1.4 GHz counts. \label{tab_counts}}
\tablewidth{0pt}
\tablehead{Flux range & Mean Flux Density &   & & & Counts &  & \\
                 & & ##Total & SF & Fraction & AGN & Fraction & RL AGN & RQ AGN \\
###$\mu$Jy & $\mu$Jy & sr$^{-1}$ Jy$^{1.5}$ & sr$^{-1}$ Jy$^{1.5}$ & \% &
  sr$^{-1}$ Jy$^{1.5}$ & \% \\
}
\startdata
$##43 - 75$ & #63 &$#2.53^{+0.51}_{-0.42}$ & $1.49^{+0.40}_{-0.32}$ &
$59^{+18}_{-17}$ & $1.03^{+0.37}_{-0.28}$ & $41^{+16}_{-14}$ & $0.16^{+0.16}_{-0.09}$ & 
$0.87^{+0.37}_{-0.27}$\\
$##75 - 120$ & #97 &$#2.62^{+ 0.45}_{-0.38}$ & $1.33^{+0.34}_{-0.28}$ &
$51^{+15}_{-14}$ & $1.24^{+0.32}_{-0.26}$ & $47^{+14}_{-13}$ & $0.60^{+0.24}_{-0.18}$  & 
$0.64^{+0.26}_{-0.19}$\\
$#120 - 200$ & 152 & $#2.87^{+0.53}_{-0.45}$ & $0.88^{+0.33}_{-0.25}$ &
$31^{+12}_{-10}$ & $1.84^{+0.44}_{-0.36}$ & $64^{+18}_{-17}$ & $1.19^{+0.37}_{-0.29}$  & 
$0.65^{+0.30}_{-0.21}$\\
$#200 - 500$ & 306 & $#4.02^{+0.74}_{-0.63}$ & $1.30^{+0.47}_{-0.36}$ &
$32^{+13}_{-11}$ & $2.61^{+0.62}_{-0.51}$ & $65^{+18}_{-17}$ & $2.11^{+0.57}_{-0.46}$  & 
$0.51^{+0.34}_{-0.22}$\\
$#500 - 2000$ & 1032 & $#6.71^{+1.86}_{-1.49}$ & $0.67^{+0.88}_{-0.43}$ &
$10^{+13}_{-7}$ & $5.70^{+1.74}_{-1.37}$ & $85^{+32}_{-31}$ & $5.37^{+1.70}_{-1.33}$  & 
$0.34^{+0.77}_{-0.28}$\\
$2000 - 100000$ & 17262 &$42.4^{+12.5}_{-9.9}$ & ... & ... ## &
$42.4^{+12.5}_{-9.9}$ & $100^{+38}_{-38}$ & $42.4^{+12.5}_{-9.9}$  & ... \\
\enddata
\end{deluxetable*}

\begin{figure}%4
%\plotone{newcounts.eps}
\centerline{\includegraphics[width=9.5cm]{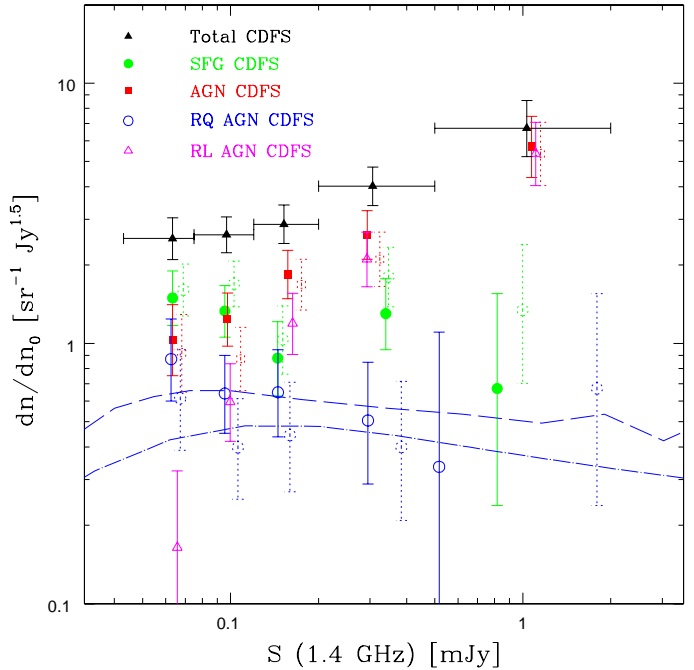}}
%\plotone{fig4.eps}
\caption{The Euclidean normalized 1.4 GHz CDFS source counts: total counts
  (black triangles), SFG (filled green circles), all AGN (red squares),
  radio-quiet AGN (open blue circles), and radio-loud AGN (open magenta
  triangles). Error bars correspond to $1\sigma$ errors \citep{geh86}. The
  dotted symbols are the source counts from Paper IV (shifted by 0.035 dex for 
  clarity). The long-dashed and
  dot-dashed lines are the radio-quiet AGN number counts predicted by
  \cite{wil08} and \cite{pad09} respectively, based on X-ray data. See text
  for details.}
\label{newcounts}
\end{figure}

Table \ref{tab_counts} and Fig. \ref{newcounts} present the Euclidean
normalized number counts for the revised sample, compared to those from
Paper IV shown in the figure. As expected from the revised classification,
SFG show a small decrease, while AGN and radio-quiet AGN increase slightly
in number. However, the revised number counts are still within $1 \sigma$
from the old ones, and most of the largest changes happen at higher flux
densities. Note that SFG are still predominant below $\sim 0.1$ mJy, 
which is also the flux density at which radio-quiet AGN start to outnumber 
radio-loud ones. 

AGN make up $49^{+7}_{-6}\%$ \citep[where the errors are based on binomial
  statistics:][]{geh86} of sub-millijansky sources and their counts are seen to
drop at lower flux densities, going from 100\% of the total at $\sim 10$
mJy down to $41\%$ at the survey limit. SFG, on the other hand, which
represent $50^{+8}_{-7}\%$ of the sample, are missing at high flux
densities but become the dominant population below $\approx 0.1$ mJy,
reaching $59\%$ at the survey limit. Radio-quiet AGN represent
$28^{+6}_{-5}\%$ (or $57\%$ of all AGN) of sub-millijansky sources but their
fraction appears to increase at lower flux densities, where they make up
$84\%$ of all AGN and $\approx 34\%$ of all sources at the survey limit, up
from $\approx 5\%$ at $\approx 1$ mJy.

\cite{mid11} have recently detected with the Very Long Baseline Array
(VLBA) 20 VLA-CDFS sources using a resolution of $\sim 0.025^{{\prime}{\prime}}$. 
With a limit of $\sim 0.5$ mJy, very long
baseline interferometry (VLBI) detections above $z > 0.1$ are most likely
to be due to AGN. Reassuringly, all of the 20 detected VLBA objects (which have 
$z > 0.15$) were classified as radio-loud AGN by our method.

\begin{figure}%5
\centerline{\includegraphics[width=9.5cm]{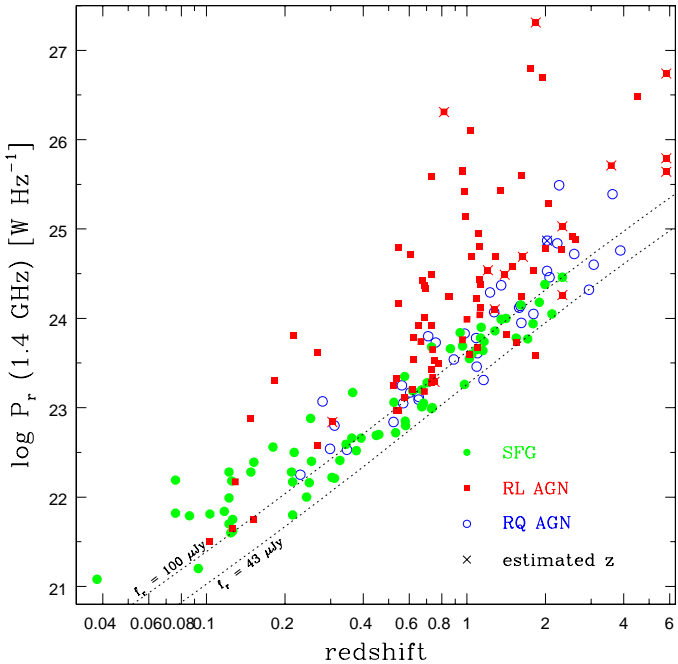}}
%\plotone{fig5.eps}
\caption{Radio power versus redshift for our sample. Filled circles
  indicate SFG, filled squares radio-loud AGN, and open circles radio-quiet
  AGN. Crosses denote redshifts estimated from the optical magnitude. The
  power - redshift relationships for a flux density of $43~\mu$Jy (lower
  dotted line) and $100~\mu$Jy (upper dotted line) assuming a radio
  spectral index of 0.7 are also shown.}
\label{prz}
\end{figure}

Fig. \ref{prz} shows radio power versus redshift for our sources, with the
dotted lines indicating $43~\mu$Jy, the faintest radio flux density of our
sample (lower line), and $100~\mu$Jy (upper line: see Section
\ref{evolution}) for $\alpha_{\rm r} = 0.7$.

\section{Evolution}\label{evolution}

\subsection{$V_{\rm e}/V_{\rm a}$ analysis}

We first study the evolutionary properties of the VLA-CDFS sample through a
variation of the $V/V_{\rm max}$ test \citep{sch68}, the $V_{\rm e}/V_{\rm
  a}$ test \citep{av80, mor91}, that is the ratio between {\it enclosed}
and {\it available} volume. This is because we do not have a single flux
limit but the sensitivity of our sample is a function of the position in
the field of view (see Paper I) and, consequently, the area of the sky
covered at any given flux density (usually known as the sky coverage) is
flux dependent. This ranges from a maximum of 0.2 deg$^2$ for radio flux
densities $\ge 295$ mJy, to 0.14 deg$^2$ at 100 mJy, down to 0.01 deg$^2$
at the flux density limit. 

Values of $\langle V_{\rm e}/V_{\rm a} \rangle$ significantly different
from 0.5 and a distribution significantly different from uniform indicate
evolution, which is positive (sources were more luminous and/or more
numerous in the past) when $\langle V_{\rm e}/V_{\rm a} \rangle > 0.5$, and
negative (sources were less luminous and/or less numerous in the past) when
$< 0.5$.  Moreover, one can also fit an evolutionary model to the sample by
finding the evolutionary parameter which makes $\langle V_{\rm e}/V_{\rm a}
\rangle = 0.5$. Note that the $V_{\rm e}/V_{\rm a}$ test is independent of
the shape of the luminosity function, unlike the maximum likelihood method
used below.

We have computed $V_{\rm e}/V_{\rm a}$ values for our sources taking into
account the appropriate sky coverage \citep[see eqs. 42 and 43 of][]{av80},
k-correcting the radio powers as described in Section~\ref{farIR}.
Statistical errors are given by
$\sigma = 1/\sqrt{12~N}$ \citep{av80}. We estimate the significance of the
deviation from the non-evolutionary case by assessing the probability
$P_{\rm ev}$ that the $V_{\rm e}/V_{\rm a}$ distribution is different from
uniform according to a Kolmogorov-Smirnov (KS) test. Similar results are
obtained by using the deviation from 0.5 of $\langle V_{\rm e}/V_{\rm a}
\rangle$. To have a simple estimate of the sample evolution we have also
derived the best fit parameter $k_{L}$ assuming a pure luminosity evolution
(PLE) of the type $P(z) = P(0) (1+z)^{k_L}$ or a pure density evolution
(PDE) of the type $\Phi(z) = \Phi(0) (1+z)^{k_D}$, where $\Phi(z)$ is the
luminosity function.

%\clearpage
\begin{deluxetable*}{lrlrcrrr}
%\begin{deluxetable}{lrlcccc}
\tabletypesize{\scriptsize}
\tablecaption{Sample Evolutionary Properties: $V_{\rm e}/V_{\rm
a}$ analysis \label{tabveva}}
\tablewidth{0pt}
\tablehead{
% ~&~&~&\multispan2{$H_0 = 70$, $\Omega_{\rm M} = 0.3$, $\Omega_{\rm
% \Lambda} = 0.7$~~~~}& \multispan2{~~~~$H_0 = 50$, $\Omega_{\rm M} = 0$,
% $\Omega_{\rm \Lambda} = 0$}\\
\colhead{Sample}&\colhead{N}&\colhead{$\langle z
\rangle$}&\colhead{estimated $z$ \%}&\colhead{$\langle V_{\rm e}/V_{\rm
a}\rangle$}&$P_{\rm ev}$&\colhead{$k_L$\tablenotemark{a}}&\colhead{$k_D$\tablenotemark{b}}
%&\colhead{$\langle V_{\rm e}/V_{\rm
%a}\rangle$}&\colhead{$\tau$}
\\}
\startdata
All sources & 193 & $1.18\pm 0.07$ & 8.3\%& $0.544\pm0.021$ & 98.8\% & 
$0.9^{+0.3}_{-0.4}$ & ...  ##\\
Star-forming galaxies & 71 & $0.90\pm 0.07$ & 1.4\% &$0.655\pm0.034$ & $>99.9\%$ &
$2.5^{+0.2}_{-0.3}$ & ...  ##\\
All AGN & 122 & $1.44\pm 0.09$ & 12.3\% & $0.479\pm0.026$ & 65.6\% &
\tablenotemark{c} ## & \tablenotemark{c} ##\\
Radio-quiet AGN & 36 & $1.73 \pm 0.16$& 2.8\%& $0.727\pm0.048$ & $>99.9\%$ &
$2.5^{+0.2}_{-0.3}$ & ...  ##\\
Radio-loud AGN & 86 & $1.26\pm 0.11$& 16.3\%& $0.375\pm0.031$ & 99.5\% &
$-3.0^{+1.0}_{-1.1}$ & $-1.6\pm0.4$\\
Radio-loud AGN, $P < 10^{24.5}$ W Hz$^{-1}$ & 53 & $0.84\pm 0.06$& 9.4\%& $0.432\pm0.040$ 
& 76.2\% & \tablenotemark{c} ## &  \tablenotemark{c} ##\\
Radio-loud AGN, $P > 10^{24.5}$ W Hz$^{-1}$ & 33 & $2.01\pm 0.26$& 27.3\%& 
$0.285\pm0.050$ & 99.9\% & ... ## & $-1.8^{+0.8}_{-0.9}$\\
\enddata
\tablenotetext{a}{Pure luminosity evolution $P(z) = P(0)(1+z)^{k_L}$}
\tablenotetext{b}{Pure density evolution $\Phi(z) = \Phi(0) (1+z)^{k_D}$}
\tablenotetext{c}{$P_{\rm ev} < 95\%$: no evolution required}
\end{deluxetable*}

We assume that some luminosity evolution takes place, based on previous
studies in the radio and other bands. When the best fit indicates negative
luminosity evolution (i.e., $k_{L} <0$), however, we fit a pure density
evolution model as well, which we feel is more physical in this case. Note
that for a single power law LF $\Phi(P) \propto P^{-\gamma}$ the
evolutionary parameters in the two cases are related through the simple
relationship $k_D = k_L(\gamma - 1)$ \citep[e.g.,][]{mar83}.

Our results are shown in Table \ref{tabveva}, which gives the sample in
column (1), the number of sources in column (2), the mean redshift in
column (3), the percentage of sources with redshift estimated from the
magnitude in column (4), $\langle V_{\rm e}/V_{\rm a} \rangle$ in column
(5), the probability $P_{\rm ev}$ that the $V_{\rm e}/V_{\rm a}$ distribution is
different from uniform in column (6), and the best fit parameters $k_L$ and
$k_D$ (when applicable) in column (7) and (8) respectively (only when
$P_{\rm ev} > 95\%$). The mean redshift is calculated taking into account
the effect of the sky coverage, that is each object is weighted by the
inverse of the area accessible at the flux density of the source
\citep[see, e.g.,][]{pad07}.

\begin{figure}%6
%\plotone{histz.eps}
\centerline{\includegraphics[width=9.5cm]{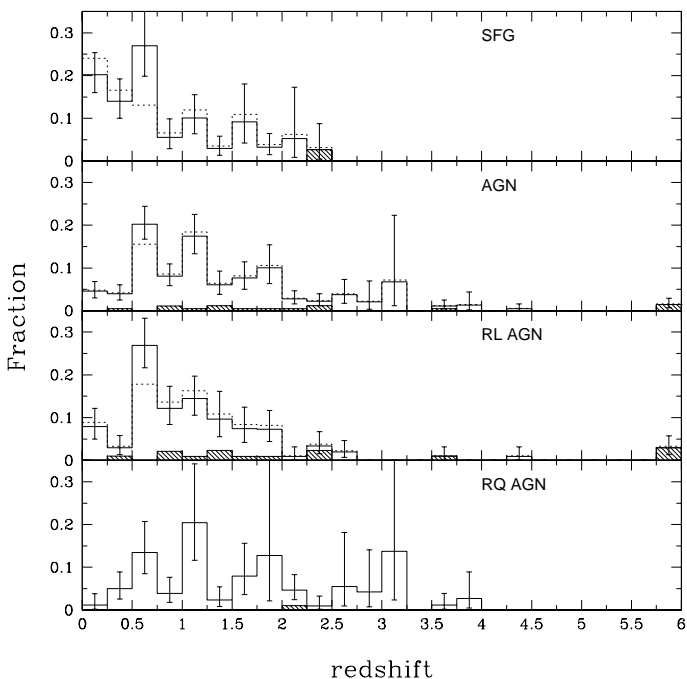}}
%\plotone{fig6.eps}
\caption{Fractional redshift distributions for the different classes of
  sources, deconvolved with the appropriate sky coverage. The dashed areas
  denote redshifts estimated from the optical magnitude. Error bars
  represent the $1\sigma$ range based on Poisson statistics. The dotted
  lines exclude all sources in the two large-scale concentrations at $0.664
  \le z \le 0.685$ and $0.725 \le z \le 0.742$ \citep{gil03}.}
\label{histz}
\end{figure}

The fractional redshift distributions for the different classes are shown
in Fig. \ref{histz}. As also shown in Table \ref{tabveva}, radio-quiet AGN
have the highest $\langle z \rangle$ and reach $z \sim 3.9$. Radio-loud AGN
have the broadest distribution, extending up to $z \sim 6$ but with a lower
$\langle z \rangle$ than radio-quiet ones. SFG, despite the relatively
narrow redshift range ($z \le 2.3$) have a $\langle z \rangle$ not too
different from that of radio-loud AGN and much smaller than radio-quiet
ones. Most classes show strong peaks at $z \sim 0.5 - 0.75$. These are
related for all classes apart from radio-quiet AGN to the two large-scale
structures detected in the CDFS by \cite{gil03} in the $0.664 \le z \le
0.685$ and $0.725 \le z \le 0.742$ ranges, which, as shown in the figure,
contribute substantially to the observed peaks. The effect of these
structures on our results is discussed in Section~\ref{lss}.

\begin{figure}%7
%\plotone{veva.eps}
\centerline{\includegraphics[width=9.5cm]{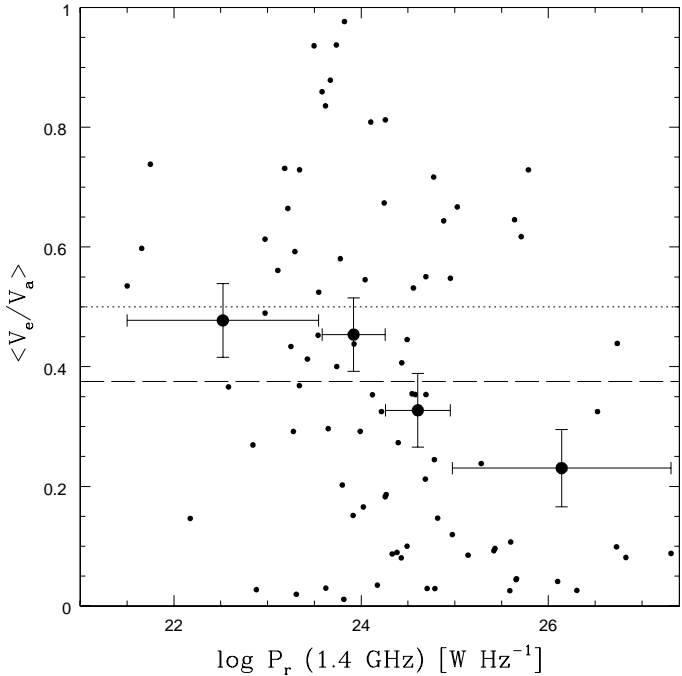}}
%\plotone{fig7.eps}
\caption{$V_{\rm e}/V_{\rm a}$ values and means versus radio power for
  radio-loud AGN. The dashed line indicates the mean value for the whole
  sample, while the dotted line is the non-evolutionary value (0.5).
  Vertical error bars represent the $1\sigma$ range based on Poisson
  statistics while the horizontal ones give the covered range of powers.}
\label{veva}
\end{figure}

The main results on the sample evolution are the following:
\begin{enumerate}

\item The whole sample has $\langle V_{\rm e}/V_{\rm a} \rangle > 0.5$ and
  shows a significant departure from the non-evolutionary case ($P_{\rm
    ev} \sim 99\%$) with an evolution characterized by
  $k_L=0.9^{+0.3}_{-0.4}$.
  
\item SFG evolve at a very high significance level ($P_{\rm ev} > 99.9\%$);
  their evolutionary parameter for the case of pure luminosity
  evolution is $k_L=2.5^{+0.2}_{-0.3}$.
 
\item AGN as a whole do not appear to evolve, as their $\langle V_{\rm e}/V_{\rm a}
  \rangle$ is only slightly below 0.5 (by $\sim 0.8 \sigma$) and the
  $V_{\rm e}/V_{\rm a}$ distribution is not significantly different from
  uniform ($P_{\rm ev} \sim 66\%$).
    
\item Radio-quiet AGN, however, evolve very significantly ($P_{\rm ev} > 99.9\%$)
  with $k_L=2.5^{+0.2}_{-0.3}$, the same value as that of SFG.
 
\item Radio-loud AGN also evolve significantly ($P_{\rm ev} = 99.5\%$) but
  in the negative sense, with $k_L=-3.0^{+1.0}_{-1.1}$ or
  $k_D=-1.6\pm0.4$. However, this is largely due to the high power
  sources. Fig. \ref{veva} shows that $V_{\rm e}/V_{\rm a}$ values are
  strongly dependent on radio power, with $\langle V_{\rm e}/V_{\rm a}
  \rangle$ becoming significantly ($\sim 2.8 \sigma$) smaller than 0.5 for
  $P > 2 \times 10^{24}$ W~Hz$^{-1}$.
  We then split the radio-loud AGN sample at $P = 10^{24.5}$
  W~Hz$^{-1}$. The low-power sub-sample does not appear to evolve, as its
  $\langle V_{\rm e}/V_{\rm a} \rangle$ is not significantly ($\sim 1.7
  \sigma$) $< 0.5$ and its $V_{\rm e}/V_{\rm a}$ distribution is not
  significantly different from uniform ($P_{\rm ev} \sim 76\%$). On the
  other hand, the high-power sub-sample anti-evolves at a very high
  significance level ($P_{\rm ev} \sim 99.9\%$), with
  $k_D=-1.8^{+0.8}_{-0.9}$. Because of the luminosity difference, the two
  sub-samples have also very different redshift distributions, with
  $\langle z \rangle \sim 0.8$ (range: $\sim 0.1- 2.3$) and 2.0 (range:
  $\sim 0.5 - 5.8$) respectively. This also means that the low-redshift
  evolution of the high-power sub-class is not well determined as, for
  example, only eight such sources ($24\%$) have $z \le 1$ (see
  Section~\ref{sect:LF}).
\end{enumerate}

\subsection{Maximum Likelihood Analysis}\label{sect:maxl}

%\clearpage
\begin{deluxetable*}{llccrc}
%\begin{deluxetable}{lrlcccc}
\tabletypesize{\scriptsize}
\tablecaption{Sample Luminosity Functions and Evolution: maximum likelihood
  analysis \label{tabmaxl}}
\tablewidth{0pt} \tablehead{
% ~&~&~&\multispan2{$H_0 = 70$, $\Omega_{\rm M} = 0.3$, $\Omega_{\rm
% \Lambda} = 0.7$~~~~}& \multispan2{~~~~$H_0 = 50$, $\Omega_{\rm M} = 0$,
% $\Omega_{\rm \Lambda} = 0$}\\
\colhead{Sample}&\colhead{Model}&\colhead{$\gamma_1$}&\colhead{$\gamma_2$}
&\colhead{$k$}&$\log P_*$\\
\colhead{}&\colhead{}&\colhead{}&\colhead{}&\colhead{}&\colhead{W~Hz$^{-1}$}}
\startdata
Star-forming galaxies & PLE & ...  & $2.56\pm0.09$ &
$2.87^{+0.06}_{-0.21}$ & ... \\
Star-forming galaxies & PLE & $1.3^{+0.5}_{-0.9}$  & $3.15^{+0.38}_{-0.27}$ &
$2.89^{+0.10}_{-0.15}$ & $21.85^{+0.22}_{-0.27}$ \\
All AGN & PLE & ... & $1.60\pm0.05$ & $-1.5\pm0.6$ & ... \\
Radio-quiet AGN & PLE & ...  & $2.6\pm0.3$ &
$2.5^{+0.4}_{-0.5}$ & ... \\
Radio-loud AGN & PLE & ... & $1.46\pm0.06$ & $-3.7^{+1.1}_{-1.6}$ & ... \\
Radio-loud AGN & PDE & ... & $1.45\pm0.06$ & $-1.8\pm0.4$ & ... \\
Radio-loud AGN, $P < 10^{24.5}$ W Hz$^{-1}$ &  PDE & ... & $1.42^{+0.14}_{-0.12}$ & 
$-1.5^{+0.9}_{-0.8}$ & ...\\
Radio-loud AGN, $P > 10^{24.5}$ W Hz$^{-1}$ &  PDE & ... & $1.46\pm0.12$ & $-1.8\pm0.6$ & ...\\
\enddata
%\tablenotetext{a}{Excluding the 7 sources without redshift}
\end{deluxetable*}

A more general 
approach to estimate the evolution, and at the same time the LF, is to
perform a maximum likelihood fit of an evolving luminosity function to the
observed distribution in luminosity and redshift. This approach makes
maximal use of the data and is free from arbitrary binning; however, unlike
the $V_{\rm e}/V_{\rm a}$ test, it is model dependent. We follow the
prescription of \cite{mar83} and minimize the following quantity

\begin{equation}
S = -2 \sum_i^N \ln [\Phi(P_i,z_i)] + 
2 \int_{P_{min}}^{P_{max}}
\int_{z_{min}}^{z_{max}}\Phi(P,z)\Omega(f(P,z)){dV\over dz}dzdP
\end{equation}

where $\Phi(P,z)$ is the luminosity function, $\Omega(f)$ is the sky
coverage, and $dV$ is the differential comoving volume. The sum is extended
over the whole sample, while the double integral is computed over the
luminosity range appropriate for the adopted evolution and over the
observed redshift range \citep[see][for more details]{mar83}. The best fit
parameters are determined by minimizing $S$ and their associated errors are
computed by varying the parameter of interest until an increment $\Delta S$
over the minimum value is obtained. $1\sigma$ errors for one parameter
correspond to $\Delta S = 1.0$ while confidence contours for 1, 2, and
$3\sigma$ levels for two interesting parameters are derived for $\Delta S =
2.3$, 6.17, and 11.8 respectively \citep{pre86}. We consider one and two
power-law LFs, that is $\Phi(P) \propto P^{-\gamma_2}$ and $\Phi(P) \propto
1/[(P/P_*)^{\gamma_1} + (P/P_*)^{\gamma_2}]$ respectively.

Our results are shown in Table \ref{tabmaxl}, which gives the sample in
column (1), the evolutionary model in column (2), the two slopes (if
applicable) of the LF in columns (3) and (4), the best-fit evolutionary
parameter in column (5), and the break power (if applicable) in column
(6). Errors are $1\sigma$ for one interesting parameter. The best-fit
evolutionary parameters agree extremely well (mostly within $1\sigma$) with
those derived through the $V_{\rm e}/V_{\rm a}$ approach.
 
\begin{figure}%8
%\plotone{maxl.eps}
\centerline{\includegraphics[width=9.5cm]{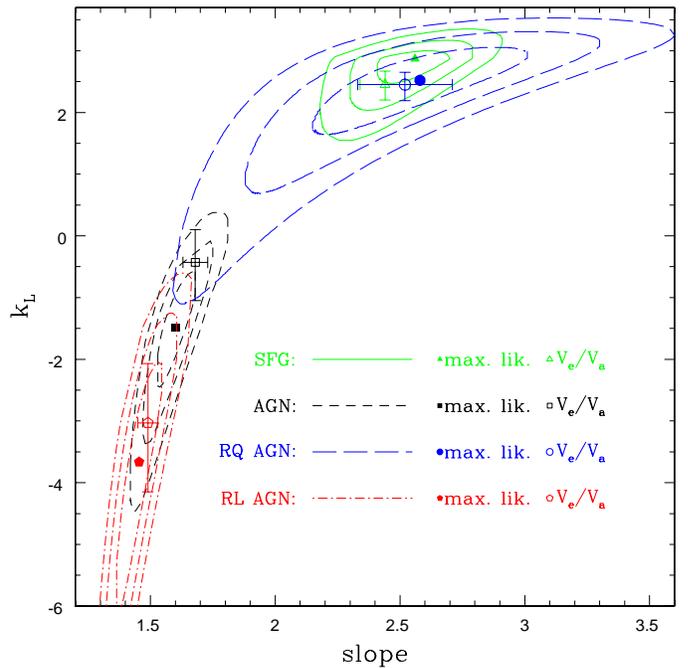}}
%\plotone{fig8.eps}
\caption{Maximum likelihood confidence contours ($1\sigma$, $2\sigma$, and
  $3\sigma$) for the evolutionary parameter $k_L$ and the single power-law
  slope of the LF. The best fit values for the various classes are indicated by
  the various symbols, with the solid ones referring to the maximum
  likelihood analysis and the open ones pertaining to the $V_{\rm e}/V_{\rm
    a}$ approach. Note that the slope of the LF in the latter case depends
  on the adopted bin size in logarithmic power and is therefore only
  indicative.}
\label{maxl}
\end{figure}

\section{Luminosity Functions}\label{sect:LF}

Fig. \ref{maxl} shows that, assuming a single power-law LF and applying the
maximum likelihood method, SFG and AGN, when considered as a single class,
have widely different forms of evolution, as seen above, but also different LF
slopes, with $\Phi(P) \propto P^{-2.6}$ and $\Phi(P) \propto P^{-1.6}$
respectively. The $3\sigma$ confidence contours have no overlap. The
situation is even more extreme when one compares SFG with radio-loud
AGN. On the contrary, in the case of the radio-quiet AGN the best fit
parameters are the same as SFG (within less than $1\sigma$) and the
confidence contours overlap to a large extent (although the radio-quiet AGN
contours are wider due to their smaller number). Finally, radio-quiet and
radio-loud AGN have very different LFs and evolution parameters, with only
a tiny overlap of the $3\sigma$ confidence contours and the former having a
much steeper LF than the latter ($\Phi(P) \propto P^{-2.6}$ vs. $\Phi(P)
\propto P^{-1.45}$).

The shape of the SFG LF is more complex than for the other classes, as the
maximum likelihood double power-law fit, given in Table \ref{tabmaxl},
excludes the case $\gamma_1 = \gamma_2$ with a significance well above
$3\sigma$, which implies that a single power-law is not a good
representation of the data. The best fit has $\Phi(P) \propto P^{-1.3}$ and
$\Phi(P) \propto P^{-3.15}$ at the faint and bright end respectively, with a
break at $P \sim 7 \times 10^{21}$ W~Hz$^{-1}$. 

\begin{figure}% 9
%\plotone{sfg_rq_rl_agn.eps}
\centerline{\includegraphics[width=9.5cm]{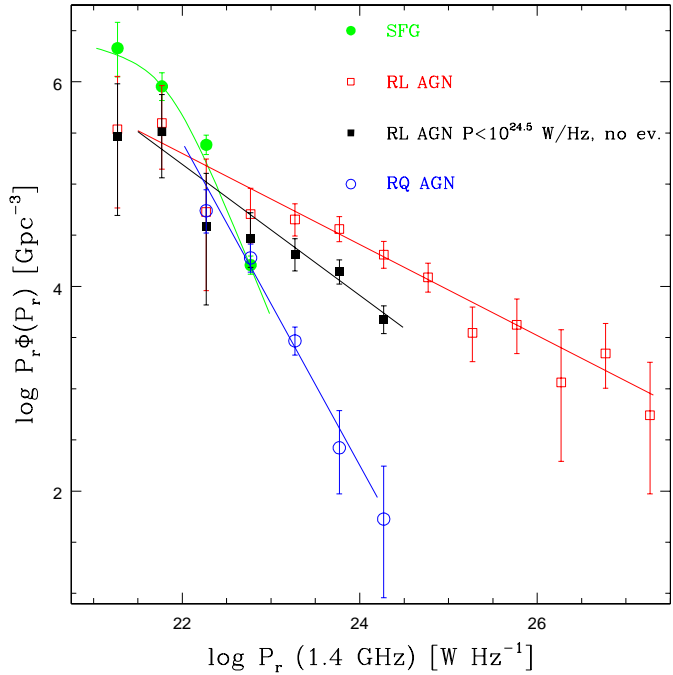}}
%\plotone{fig9.eps}
\caption{The local differential 1.4 GHz LFs for SFG and AGN in a $P \times \phi(P)$
  form obtained with the maximum likelihood method (solid lines) and the LFs
  de-evolved to $z=0$ using the best fit evolutionary parameters from 
  the maximum likelihood analysis (points). 
  Filled circles indicate SFG, open squares radio-loud AGN, filled squares 
  radio-loud AGN with $P < 10^{24.5}$ W Hz$^{-1}$, and
  open circles radio-quiet AGN. The LFs for radio-loud AGN with $P < 
  10^{24.5}$ W Hz$^{-1}$ from both methods are shown without any
  de-evolution as inferred from the $\langle V_{\rm e}/V_{\rm a} \rangle$
  analysis. Error bars correspond to $1\sigma$ Poisson errors
  \citep{geh86} evaluated using the number of sources per bin with redshift
  determination only. See text for details.}
\label{sfg_rq_rl}
\end{figure}

Fig. \ref{sfg_rq_rl} shows the local LFs derived from the the maximum
likelihood best fits compared with those obtained from the $1/V_{\rm max}$
(in our case $1/V_{\rm a}$) technique without any assumption on the LF but
de-evolved to $z=0$ using the best fit evolutionary parameters. The LFs
are shown in a $P \times \Phi(P)$ form, which is almost
equivalent\footnote{$P \times \Phi(P) = 2.5/ln(10) \times \Phi(M)
  \sim 1.09 \times \Phi(M)$, where the units of $\Phi(M)$ are mag$^{-1}$ 
  Volume$^{-1}$. Note that these units are also
  sometimes used in the radio band: e.g., \cite{con89,sad02,mau07}.}  to the
$\phi(M_{\rm B})$ form normally used in the optical band and allows an easy
separation of luminosity and density evolution as the former would simply
translate the LF to the right (higher powers) with no change in the
ordinate (number), while the opposite would be true for the latter.

The maximum likelihood fits, although relatively simple, appear to be very
good representations of the LFs obtained with the $1/V_{\rm a}$ technique.
From Fig. \ref{sfg_rq_rl} one also infers that AGN dominate over SFG for $P
\ga 3 \times 10^{22}$ W~Hz$^{-1}$, in agreement with previous studies
\citep[e.g.,][]{mau07}. Moreover, radio-loud AGN have a much flatter LF
than radio-quiet ones and are predominant at $P \ga 6 \times 10^{22}$
W~Hz$^{-1}$. Finally, the radio-quiet AGN LF seems to be an extension of
the SFG one at higher radio powers.

It is important to be aware of the fact that the low-redshift behavior of
the high-power radio-loud AGN is not well determined. Based on previous
results (see Section~\ref{sect:rlagn}) we would in fact expect a strong
positive evolution at moderately low redshifts followed by a decline at
higher redshifts. Our sample is too small to detect such a change, since it
is dominated by high redshift objects ($\sim 3/4$ of the sample has $z >
1$). This means that the best fit PDE reflects largely the high redshift
negative evolution and, once the LF is de-evolved to $z = 0$, this
translates into an artificially large density of sources at high powers. We
therefore plot in Fig. \ref{sfg_rq_rl} also the LF of $P < 10^{24.5}$ W
Hz$^{-1}$ sources with their evolution fixed to zero, based on the results
of the $\langle V_{\rm e}/V_{\rm a} \rangle$ analysis, which should give a
more realistic estimate of the local LF.

We now concentrate on the details of the individual classes. 

\subsection{Star-forming galaxies}

\begin{figure}%10
%\plotone{sfg_local.eps}
\centerline{\includegraphics[width=9.5cm]{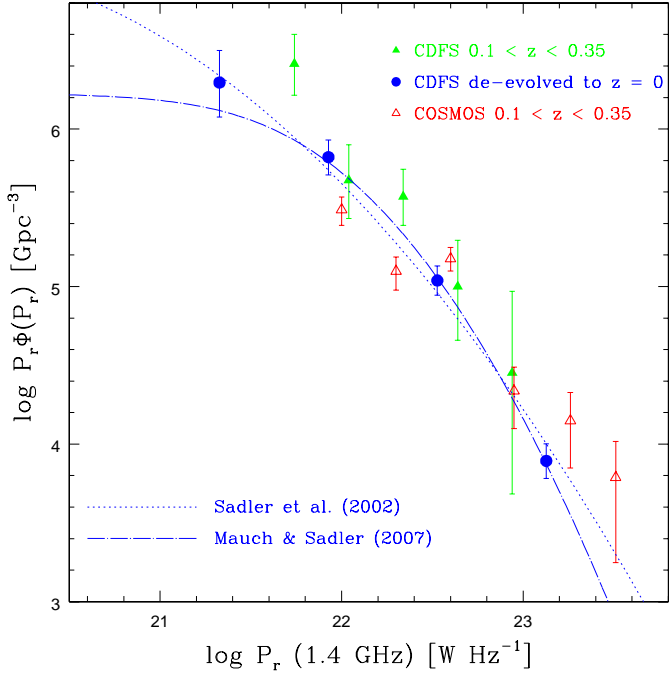}}
%\plotone{fig10.eps}
\caption{The local differential 1.4 GHz LF for SFG in a $P \times \phi(P)$
  form. Filled triangles indicate the VLA-CDFS LF in the $0.1 - 0.35$
  redshift range, open triangles refer to the COSMOS LF in the same range,
  and filled circles denote the LF de-evolved to $z=0$ using the best fit
  evolutionary parameter from the $\langle V_{\rm e}/V_{\rm a} \rangle$
  analysis. The best fits to the local SFG LF from
  \cite{sad02} (converted to our value of $H_0$) and \cite{mau07} are also
  shown (dotted and dash-dotted lines respectively). Error bars correspond
  to $1\sigma$ Poisson errors \citep{geh86} evaluated using the number of
  sources per bin with redshift determination only. See text for details.}
\label{sfg_local}
\end{figure}

Fig. \ref{sfg_local} shows different estimates of the local LF for our SFG.
Filled triangles indicate the VLA-CDFS SFG LF in the $0.1 - 0.35$ redshift
range, to compare it with the COSMOS LF \citep{smo09a} in the same range
(open triangles). The two LFs are within $1\sigma$ apart from one bin. Our
LF is ``noisier'' as we only have 24 objects in this redshift bin, as
compared to 98 for the COSMOS sample. We also note that the selection
criteria of the two samples are very different, with the COSMOS one being
based on a rest-frame optical color classification \citep{smo08}.

\begin{deluxetable}{lclr}
%\begin{deluxetable}{lrlcccc}
\tabletypesize{\scriptsize}
\tablecaption{Luminosity Functions for VLA-CDFS star-forming galaxies\label{tab:sfg_lf}}
\tablewidth{0pt} \tablehead{
\colhead{Redshift range}&\colhead{$\log P_{\rm 1.4~GHz}$}&\colhead{$\log P \Phi(P)$}&
\colhead{~~N}\\
\colhead{}&\colhead{W~Hz$^{-1}$}&\colhead{Gpc$^{-3}$}&\colhead{}
}
\startdata
  & 21.27 & $6.39^{+0.36}_{-0.45}$ &    2 \\   
$0.038 < z \le 0.214$ &21.67 & $6.22^{+0.16}_{-0.17}$ &    9 \\   
  &22.07 & $5.55^{+0.25}_{-0.28}$ &    4 \\  
  &22.47 & $5.61^{+0.22}_{-0.24}$ &    5 \\  
\hline
 &22.19 & $5.84^{+0.25}_{-0.28}$ &    4 \\   
$0.214 < z \le 0.524$ &22.59 & $5.33^{+0.16}_{-0.17}$ &    9 \\   
 &22.99 & $4.40^{+0.29}_{-0.34}$ &    3 \\   
 \hline
 &22.91 & $5.54^{+0.17}_{-0.18}$ ($5.44^{+0.29}_{-0.34}$) &    8 (3) \\   
$0.524 < z \le 1.021$  &23.31 & $4.36^{+0.22}_{-0.24}$ ($4.22^{+0.25}_{-0.28}$) &    5 (4) \\   
 &23.71 & $4.03^{+0.22}_{-0.24}$ ($3.90^{+0.25}_{-0.28}$) &    5 (4) \\   
\hline
 &23.79 & $5.03^{+0.15}_{-0.16}$ &   10 \\   
$1.021 < z \le 2.325$  &24.19 & $3.70^{+0.20}_{-0.22}$ &    6 \\   
 &24.59 & $2.71^{+0.52}_{-0.77}$ &    1 \\   
\enddata
\tablenotetext{~}{Numbers in parenthesis exclude the seven sources
  in the two large-scale concentrations at $0.664 \le z \le 0.685$ and
  $0.725 \le z \le 0.742$ \citep{gil03}. Errors correspond
  to $1\sigma$ Poisson errors \citep{geh86} evaluated using the number of
  sources per bin with redshift determination only. The conversion to units of
  Mpc$^{-3}$ dex$^{-1}$ used, for example, by \cite{smo09a}, is done by
  subtracting $9 - \log(\ln(10))$ from our values.}
\end{deluxetable}

\begin{figure}%11
%\plotone{sfg.eps}
\centerline{\includegraphics[width=9.5cm]{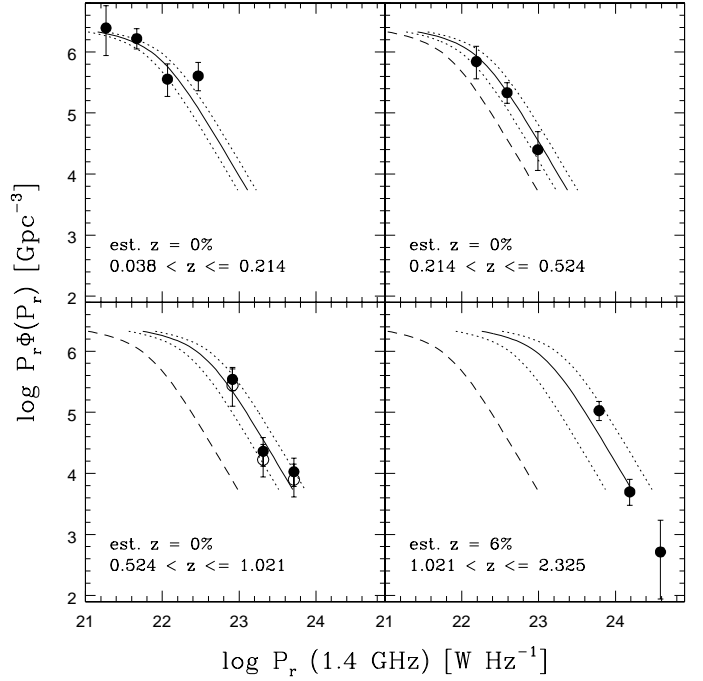}}
%\plotone{fig11.eps}
\caption{The differential 1.4 GHz LF for VLA-CDFS SFG in a $P \times
  \phi(P)$ form in four redshift bins. The solid lines represent the best
  fit double power-law LF from the maximum likelihood method evolved to the
  central redshift of the bin using the best-fit evolution $(1+z)^{2.89}$,
  with dotted lines showing the same LF at the two extreme redshifts
  defining the bin. The short-dashed line represents the LF at $z=0$. Error
  bars correspond to $1\sigma$ Poisson errors \citep{geh86} evaluated using
  the number of sources per bin with redshift determination only. The
  percentage of redshifts estimated from the optical magnitude is also
  given for each bin. Open symbols in the $0.524 - 1.021$ bin do not
  include the seven sources in the two large-scale concentrations at $0.664
  \le z \le 0.685$ and $0.725 \le z \le 0.742$ \citep{gil03}. See text for
  details.}
\label{sfg}
\end{figure}

To have a more robust estimate of the local LF for SFG we have derived the
LF de-evolved to $z=0$ (filled circles) using the best fit evolutionary
parameter from the $\langle V_{\rm e}/V_{\rm a} \rangle$ analysis. 
This makes use of the whole sample but is obviously dependent on the
assumed evolutionary model. In Fig. \ref{sfg_local} we also plot the best
fits to the local ($z \le 0.3$) LFs from \cite{sad02} and \cite{mau07}
(dotted and dash-dotted lines respectively). Both our local LFs are
consistent with these two LFs, particularly the de-evolved one
($\chi^2_{\nu} < 0.6$). This validates our selection method.  As was the
case for the maximum likelihood approach, a single power-law fit is
inconsistent with the de-evolved local LF ($\chi^2_{\nu} \sim 3.4$,
significant at the $\sim 97\%$ level).

The maximum likelihood fit provides a very good representation of the redshift 
evolution of the LF of SFG, as shown in Fig. \ref{sfg} 
(tabulated in Tab. \ref{tab:sfg_lf}), which plots the SFG LF
over the full redshift range sampled, that is $0 - 2.3$, in four redshift
bins each containing a similar number of objects. The median enclosed volumes 
for the four bins are $3 \times 10^{-6}, 2 \times 10^{-5}, 9 \times
10^{-5}$ and $5 \times 10^{-4}$ Gpc$^3$ respectively.

\subsection{AGN}\label{agn:LF}

\begin{figure}%12
%\plotone{agn_local.eps}
\centerline{\includegraphics[width=9.5cm]{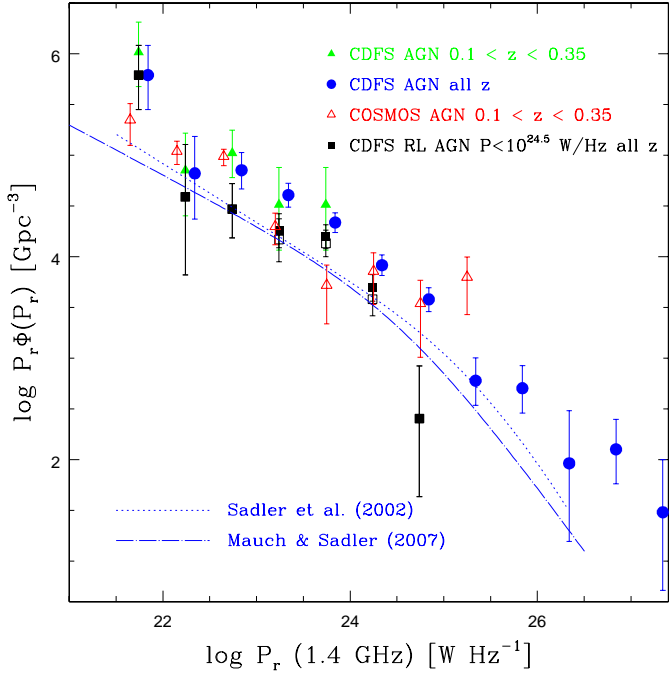}}
%\plotone{fig12.eps}
\caption{The local differential 1.4 GHz LF for AGN in a $P \times \phi(P)$
  form. Filled triangles indicate the VLA-CDFS LF in the $0.1 - 0.35$
  redshift range, open triangles refer to the COSMOS LF in the same range,
  filled circles denote the full AGN LF (shifted by 0.1 in log P for clarity), while
  filled squares are the radio-loud AGN with $P < 10^{24.5}$ W Hz$^{-1}$
  (with open squares showing the effect of excluding the ten sources in
  the two large-scale concentrations at $0.664 \le z \le 0.685$ and $0.725
  \le z \le 0.742$: \cite{gil03}). Both of these LFs are shown without any
  de-evolution as inferred from the $\langle V_{\rm e}/V_{\rm a} \rangle$
  analysis. The best fits to the local AGN LF from \cite{sad02} (converted
  to our value of $H_0$) and \cite{mau07} are also shown (dotted and
  dash-dotted lines respectively). Error bars correspond to $1\sigma$
  Poisson errors \citep{geh86} evaluated using the number of sources per
  bin with redshift determination only. See text for details.}
\label{agn_local}
\end{figure}

Fig. \ref{agn_local} shows different estimates of the local LF for our AGN.
Filled triangles indicate the VLA-CDFS AGN LF in the $0.1 - 0.35$ redshift
range, to compare it with the COSMOS LF \citep{smo09b} in the same range
(open triangles). The two LFs are within $\approx 1\sigma$. As was the case
for SFG, our LF is ``noisier'' as we only have 15 objects in this redshift
bin. The selection criteria for the two samples are also quite different
and quasars ($\la 20\%$ of the AGN sample) were not included by 
\cite{smo09b}. We also show the full LF for all AGN (filled circles) assuming 
no evolution, as inferred from our analysis. 

%Since the best fit evolution for radio-loud AGN is instead negative, these
%points are below the radio-loud AGN LF shown in Fig. \ref{sfg_rq_rl} for $P
%\ga 10^{24.5}$ W Hz$^{-1}$, although they are consistent with it above $P
%\approx 10^{26}$ W Hz$^{-1}$ due to small number statistics.

In Fig. \ref{agn_local} we also plot the best fits to the local ($z \le
0.3$) LFs from \cite{sad02} and \cite{mau07} (dotted and dash-dotted lines
respectively). Both of our LFs are inconsistent with these previous
estimates ($\chi^2_{\nu} \ge 3.0$, significant at the $> 98.9\%$ level) and
about a factor $\sim 2 - 4$ higher. This is also true of the COSMOS LF of
\cite{smo09b} ($\chi^2_{\nu} \sim 4.6$, significant at the $> 99.9\%$
level), despite their claims to the contrary. Two effects are at play here.
The first, and more important one, is that 
the AGN LF includes a sizable
contribution from radio-quiet AGN, which were not present in significant
numbers in previous LFs as these were based on the NVSS survey
($S_{\rm min} \ga 2.8$ mJy), while radio-quiet AGN make up a non-negligible
fraction of radio sources only below $\approx 1$ mJy
(Fig. \ref{newcounts}), and both included only non-stellar optical sources.
Note that none of these arguments apply to star-forming galaxies, which
look non-stellar and, despite their relatively low fraction at high radio
flux densities, were selected for because of the low redshift cuts (AGN
tend to have higher redshifts: see Fig. \ref{histz}) and, in the case of
the \cite{mau07} sample, the K-band selection.  (This bias is vividly
illustrated by Fig. 5 of \citealt{mau07}, where star-forming galaxies
dominate below $\sim 10$ mJy, while Fig. \ref{newcounts} shows that in
purely radio selected samples without any redshift cut this happens below
$\sim 0.1$ mJy.) If one considers only radio-loud AGN with $P < 10^{24.5}$
W Hz$^{-1}$ (filled squares), which are non-evolving, our LF is
marginally consistent with that from \cite{sad02} ($\chi^2_{\nu} \sim 2.0$,
significant at the $\sim 95\%$ level). The second effect is related to cosmic variance:
the exclusion of the eleven sources
in the two large-scale concentrations at $0.664 \le z \le 0.685$ and $0.725
\le z \le 0.742$ \citep{gil03} (open squares), in fact, reduces even further the
discrepancy with both previously determined local LFs, which are now
consistent with ours ($\chi^2_{\nu} \le 1.8$, significant at the $\le 91\%$
level).

%Indeed, if one splits the sample in two equally populated redshift bins,
%one gets that for $z \le 0.734$ $\langle V_{\rm e}/V_{\rm a} \rangle =
%0.62\pm0.05$, with $P_{\rm ev} \sim 97.5\%$, while for higher redshifts
%$\langle V_{\rm e}/V_{\rm a} \rangle = 0.30\pm0.04$ with $P_{\rm ev} >
%99.9\%$. If one excludes the sources in the two large-scale structures
%at $z \sim 0.67$ and 0.73 \citep[][see also Section~\ref{lss}]{gil03}, then 
%the lower redshift bin is consistent with no evolution ($\langle 
%V_{\rm e}/V_{\rm a} \rangle = 0.52\pm0.05$, with 
%$P_{\rm ev} \sim 52.9\%$). 

\begin{deluxetable}{lclr}
%\begin{deluxetable}{lrlcccc}
\tabletypesize{\scriptsize}
\tablecaption{Luminosity Functions for VLA-CDFS radio-loud AGN\label{tab:rlagn_lf}}
\tablewidth{0pt} \tablehead{
\colhead{Redshift range}&\colhead{$\log P_{\rm 1.4~GHz}$}&\colhead{$\log P \Phi(P)$}&
\colhead{~~N}\\
\colhead{}&\colhead{W~Hz$^{-1}$}&\colhead{Gpc$^{-3}$}&\colhead{}
}
\startdata
 &21.74 & $5.79^{+0.29}_{-0.34}$ &    3 \\   
 &22.24 & $4.59^{+0.52}_{-0.77}$ &    1 \\   
 &22.74 & $4.49^{+0.25}_{-0.28}$ &    5 \\   
$0.103 < z \le 0.651$ &23.24 & $4.22^{+0.22}_{-0.24}$ &    5 \\   
 &23.74 & $4.20^{+0.22}_{-0.24}$ &    5 \\   
 &24.24 & $3.50^{+0.52}_{-0.77}$ &    1 \\   
 &24.74 & $3.80^{+0.36}_{-0.45}$ &    2 \\   
\hline
 &23.42 & $4.53^{+0.19}_{-0.20}$ ($4.35^{+0.29}_{-0.34}$) &    8 (4) \\   
 &23.92 & $3.96^{+0.25}_{-0.28}$ ($3.34^{+0.52}_{-0.77}$) &    4 (1)\\   
$0.651 < z \le 0.964$ &24.42 & $4.11^{+0.20}_{-0.22}$ ($3.81^{+0.29}_{-0.34}$) &    6 (3)\\   
 &25.42 & $3.81^{+0.29}_{-0.34}$ ($3.63^{+0.36}_{-0.45}$)  & 3 (2)\\   
 &26.42 & $3.33^{+0.52}_{-0.77}$ ($3.33^{+0.52}_{-0.77}$) &    1 (1)\\   
\hline
 &23.85 & $4.13^{+0.20}_{-0.22}$ &    6 \\  
 &24.35 & $3.83^{+0.22}_{-0.24}$ &    8 \\   
$0.964 < z \le 1.546$  &24.85 & $3.49^{+0.25}_{-0.28}$ &    4 \\   
 &25.35 & $3.36^{+0.29}_{-0.34}$ &    3 \\   
 &25.85 & $2.89^{+0.52}_{-0.77}$ &    1 \\   
\hline
 &23.82 & $3.17^{+0.52}_{-0.77}$ &    1 \\  
 &24.32 & $3.21^{+0.36}_{-0.45}$ &    3 \\   
 &24.82 & $3.07^{+0.25}_{-0.28}$ &    6 \\   
$1.546 < z \le 5.818$  &25.32 & $1.98^{+0.52}_{-0.77}$ &    1 \\   
 &25.82 & $2.58^{+0.52}_{-0.77}$ &    4 \\   
 &26.32 & $1.96^{+0.52}_{-0.77}$ &    1 \\   
 &26.82 & $2.44^{+0.36}_{-0.45}$ &    3 \\   
 &27.32 & $1.96^{+0.52}_{-0.77}$ &    1 \\    
\enddata
\tablenotetext{~}{Numbers in parenthesis exclude the eleven sources
  in the two large-scale concentrations at $0.664 \le z \le 0.685$ and
  $0.725 \le z \le 0.742$ \citep{gil03}. Errors correspond
  to $1\sigma$ Poisson errors \citep{geh86} evaluated using the number of
  sources per bin with redshift determination only. The conversion to units of
  Mpc$^{-3}$ dex$^{-1}$ used, for example, by \cite{smo09b}, is done by
  subtracting $9 - \log(\ln(10))$ from our values.}
\end{deluxetable}

\begin{figure}%13
%\plotone{rlagn.eps}
\centerline{\includegraphics[width=9.5cm]{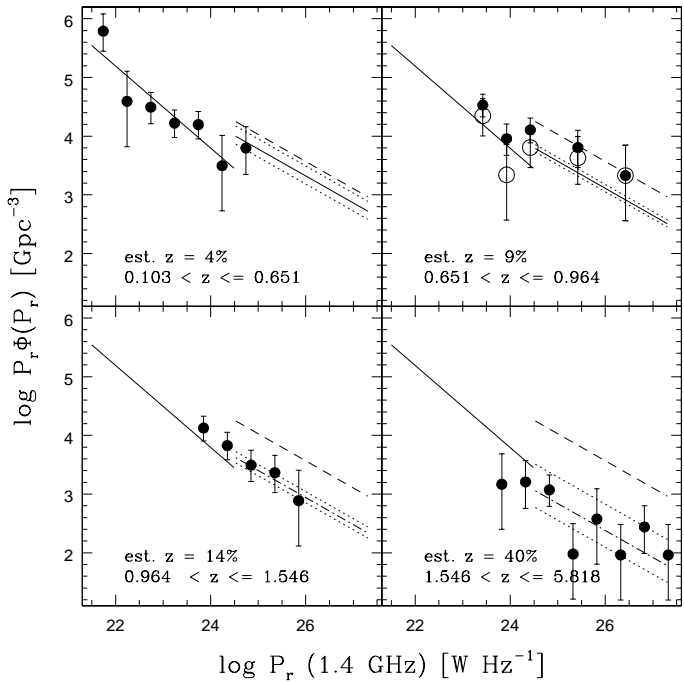}}
%\plotone{fig13.eps}
\caption{The differential 1.4 GHz LF for VLA-CDFS radio-loud AGN in a $P
  \times \phi(P)$ form in four redshift bins. The solid lines represent the
  best fit single power-law LF from the maximum likelihood method for $P <
  10^{24.5}$ W Hz$^{-1}$ (left) and $P > 10^{24.5}$ W Hz$^{-1}$ (right),
  the latter evolved to the central redshift of the bin using the best fit
  for pure density evolution $(1+z)^{-1.8}$, with dotted lines showing the
  same LF at the two extreme redshifts defining the bin. The short-dashed
  line represents the best fit LF at $z=0$ for $P > 10^{24.5}$ W Hz$^{-1}$.
  Error bars correspond to $1\sigma$
  Poisson errors \citep{geh86} evaluated using the number of sources per
  bin with redshift determination only. The percentage of redshifts
  estimated from the optical magnitude is also given for each bin. Open
  symbols in the $0.651 - 0.964$ bin do not include the eleven sources in
  the two large-scale concentrations at $0.664 \le z \le 0.685$ and $0.725
  \le z \le 0.742$ \citep{gil03}. See text for details.}
\label{rlagn}
\end{figure}

The maximum likelihood best fits for the two classes of radio-loud AGN,
below and above $P = 10^{24.5}$ W Hz$^{-1}$, provide a good representation
of the evolution and LF of radio-loud AGN, as shown in Fig. \ref{rlagn} 
(tabulated in Tab.  \ref{tab:rlagn_lf}),
which plots the radio-loud AGN LF over the full redshift range sampled, that is
$0.1 - 5.8$, in four redshift bins each containing a similar number of
objects. (For the radio-loud AGN with powers below $10^{24.5}$ W Hz$^{-1}$
we fixed the evolution to zero, based on the results of the $V_{\rm
  e}/V_{\rm a}$ analysis, excluded the sources in the two large-scale
redshift concentrations, and obtained $\Phi(P) \propto P^{-1.7\pm0.1}$).
The median enclosed volumes for the four bins are $2 \times 10^{-4}, 7 \times 
10^{-5}, 2 \times 10^{-4}$ and $10^{-3}$ Gpc$^3$ respectively. 

We note that the high-power end of the radio luminosity function is poorly
sampled because of the relatively small volume covered by our small-area
field. This, combined with the fact that $\sim 30\%$ of the redshifts for
our high-power radio-loud AGN have been estimated from the optical
magnitude, makes our results for the $P > 10^{24.5}$ W Hz$^{-1}$ sub-class
more uncertain than for the other classes. Moreover, as discussed above,
their low redshift evolution is not well determined (the two lowest
redshift bins include two and four objects respectively), which also means
that the best-fit LF derived by the maximum likelihood method 
is artificially high at $z \la 1$.

\section{Discussion}

\subsection{Effect of missing redshifts on our results}\label{missingz}

Since 14/16 of the sources without redshift information are classified as
radio-loud AGN (mostly based on their $q$ and $R$ values), the other
classes are basically unaffected by our redshift incompleteness. We checked
how the redshifts estimated from the optical magnitude influence our
results in two different ways. First, we assumed $z = \langle z \rangle =
1.13$ for these 16 sources. This is very different from our previous
assumption, which resulted in a very broad redshift distribution extending
between 0.3 and 5.8. The $\langle V_{\rm e}/V_{\rm a} \rangle$ (and best
fit evolutionary parameters) changed negligibly for all but one sub-class,
at most by an amount equal to $0.4\sigma$. Given the smaller redshifts (and
therefore luminosity) involved, in fact, the sample of radio-loud AGN with $P >
10^{24.5}$ W Hz$^{-1}$ shrank by $\sim 18\%$, while their $\langle V_{\rm
  e}/V_{\rm a} \rangle$ decreased even further by $1.5\sigma$ reaching
$\sim 0.21$. Second, we estimated the missing redshifts using the scatter
of the correlation between $\log z$ and $V_{\rm mag}$ (dashed lines in
Fig. \ref{zvmag}). The anti-evolution of radio-loud AGN was confirmed
despite the substantial redshift variations ($P_{\rm ev} > 97.2\%$), with
changes in $\langle V_{\rm e}/V_{\rm a} \rangle$ only up to $0.6
\sigma$. In the case of high-power radio-loud AGN the changes were somewhat
larger, with an increase in $\langle V_{\rm e}/V_{\rm a} \rangle$ of $1.6 \sigma$ 
when the upward scatter was
applied (which however implies estimated redshifts up to $\sim 10$). The evolution was
nevertheless still strongly negative ($P_{\rm ev} = 99\%$). As for
low-power radio-loud AGN, the changes in $\langle V_{\rm e}/V_{\rm a}
\rangle$ were $< 0.3 \sigma$ with still no significant evidence for a
departure from the non-evolutionary case ($P_{\rm ev} < 86\%$).

In summary, our results are quite insensitive to the specific redshift
values for the fraction of the sample without redshift information. This is
not surprising, as redshift affects $V_{\rm e}/V_{\rm a}$ values much less
than flux density and our redshift incompleteness is very small ($\sim
8\%$).

\subsection{Effect of large-scale structures on our results}\label{lss}

\cite{gil03} have studied the large-scale structure in the CDFS in the
X-ray and near-IR bands and detected two concentrations of sources in the
$0.664 \le z \le 0.685$ and $0.725 \le z \le 0.742$ ranges. Given the small
area of our survey one could worry that such redshift spikes might
influence some of our results. Indeed, the redshift distributions shown in
Fig. \ref{histz} peak in the $0.5 - 0.75$ bin for most classes but the
peaks become much less pronounced when these objects are excluded. There
are in fact 18 sources in these two redshift bins (7 SFG and 11 AGN), which
make up $\sim 17\%$ and $\sim 11\%$ (taking into account the effect of the
sky coverage) of all SFG and radio-loud AGN respectively (but 0\% of
radio-quiet AGN and only $\sim 3\%$ of high-power radio-loud AGN).

To assess the {\it maximum} impact of these two structures on our results we
studied the evolution of our sources by excluding {\it all} sources in
these two redshift bins. The resulting $\langle V_{\rm e}/V_{\rm a}
\rangle$ values and best-fit evolutionary parameters were within $1 \sigma$
from those derived from the full samples for all classes, which shows that
the effect of these large-scale structures on our results is minimal. These
over-densities are obviously more noticeable when one studies the evolution
of the LF with redshift (see, e.g., the bottom-left panel of Fig.
\ref{sfg} and the top-right panel of Fig. \ref{rlagn}) but even then the
revised LFs are within $\la 1\sigma$ from the old ones. The exclusion
of these sources has also some effect when we limit the SFG and radio-loud AGN
samples to $z \le 1.3$ (Section~\ref{comp_COSMOS}).

\subsection{Are there two classes of low radio power AGN?}\label{2_lowrl}

\begin{figure}%14
%\plotone{l_q.eps}
\centerline{\includegraphics[width=9.5cm]{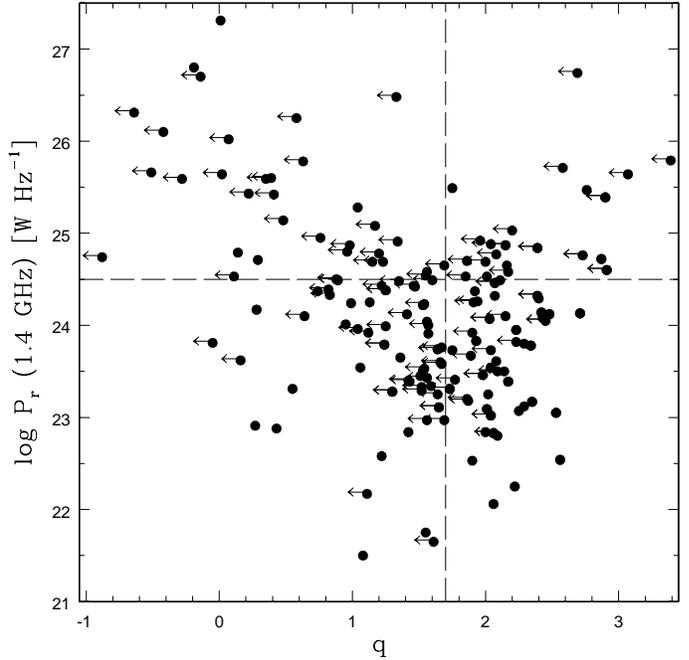}}
%\plotone{fig14.eps}
\caption{Radio power versus $q$, the logarithm of the ratio between far-IR
  and radio powers for AGN in our sample. The vertical dashed line
  indicates $q = 1.7$, the dividing value between radio-quiet and
  radio-loud AGN, while the horizontal dashed line ($P_{\rm r} = 10^{24.5}$
  W Hz$^{-1}$) divides low- and high-power radio-loud AGN. See text for
  details.}
\label{l_r_q}
\end{figure}

We have identified two classes of low-power AGN: radio-quiet ones, defined
as spelled out in Section~\ref{sample}, $\sim 94\%$ of which turn out to have 
$P_{\rm r} < 10^{25}$ W Hz$^{-1}$, and radio-loud ones, characterized by
$P_{\rm r} \la 3 \times 10^{24}$ W Hz$^{-1}$. Both classes have also
relatively low $R$ values, as implicit in our selection of radio-quiet sources 
and as shown in Fig. 3 of \cite{pad09}. One obvious question is
then what the differences between these two classes are. The answer is:
many. First, they have very different ($P > 99.99\%$) distributions in IRAC
flux ratios (Section~\ref{sample}), according to a two-dimensional KS test
\citep{fa87}, with radio-loud AGN mostly towards the old stellar population
locus and the majority of radio-quiet AGN populating the region where most
unobscured AGN should be (note that none of the selection requirements for
the radio-loud/radio-quiet distinction was based on that). Second, despite
the similar radio powers, they have very different redshift distributions
($P > 99.9\%$) and $\langle z \rangle$, 1.73 and 0.84 for radio-quiet and
low-power radio-loud AGN respectively. Third, they evolve very differently
and have different LFs, with radio-quiet AGN characterized by a steep LF and
strong positive evolution while low-power radio-loud AGN display a much
flatter LF and no evolution (and in any case $\langle V_{\rm e}/V_{\rm a}
\rangle < 0.5$; see Section~\ref{evolution}). Fourth, they have very
different X-ray-to-radio luminosity distributions and ratios. To properly
take into account the upper limits on X-ray power we used ASURV
\citep{la92}, the Survival Analysis package which employs the routines
described in \cite{fei85} and \cite{iso86} and evaluates differences in
distributions and mean values by dealing appropriately with
non-detections. The distributions for the two classes are significantly
different ($P > 99.95\%$), with radio-quiet AGN having an order of
magnitude larger X-ray power for the same radio power. Fifth, for the
sources for which we have a spectrum the optical classification is very
different: $\sim 2/3$ of the radio-quiet AGN with a spectrum in
\cite{szo04} are either broad-lined or high-excitation line objects, while
only $\sim 8\%$ of the low-power radio-loud AGN are classified as such (one
object with $P_{\rm r} \sim 1.5 \times 10^{24}$ W Hz$^{-1}$, which is close 
to the dividing line with radio powerful AGN). Finally, even
though the definition of the two classes is based on their $q$ values, the
point remains that low radio power AGN span $\sim 3$ orders of magnitude in
far-IR to radio flux density ratios (see Fig. \ref{l_r_q}), meaning that
their radio emission goes from being related to star formation to having a
likely jet origin. Fig. \ref{l_r_q} shows also that this is not the case
for high-power ($P_{\rm r} \ga 3 \times 10^{24}$ W Hz$^{-1}$) AGN, since
they basically all have $q \la 1.7$ or upper limits above this value.

\subsection{Are our radio-quiet AGN and star-forming galaxies really 
different classes?} 

We have also identified two classes of radio sources with $q$ values
suggesting a star-formation origin for their radio emission. Both classes
have low $R$ and low $P_{\rm r}$ (SFG by definition). They also evolve 
similarly in the radio band, have a similar slope of the LF, and the same range
of $q$. Are we really dealing with two different classes? Again, the answer
is: yes. First, they have very different ($P > 99.99\%$) distributions in
IRAC flux ratios (Section~\ref{sample}), according to a two-dimensional KS
test, even before the few outliers were removed ($P \sim 99\%$), with
radio-quiet AGN populating the region where AGN should be and SFG
distributed in the region where PAH- and starlight-dominated sources are
expected to be. Second, despite their similar radio powers, they have very
different redshift distributions ($P > 99.99\%$) and $\langle z \rangle$,
1.73 and 0.90 for radio-quiet AGN and SFG respectively. Third, for the
sources for which we have a spectrum, the optical classification is very
different: $\sim 2/3$ of the radio-quiet AGN with a spectrum in
\cite{szo04} are either broad-lined or high-excitation line objects, while
none of the SFG are classified as such. Finally, even though the definition
of the two classes is based also on their X-ray powers, the point remains
that radio sources with $q \ge 1.7$ span almost six orders of magnitude in
$L_{\rm x}$ ($\sim$ three orders of magnitude on either side of the
$10^{42}$ ergs s$^{-1}$ divide). This means that their X-ray emission goes
from being related to star formation to having a clear AGN origin. In
summary, although radio-quiet AGN and SFG radio have similar $q$ values and
other radio-related properties, in one case we are clearly dealing with an
AGN while in the other we are not.

\subsection{The evolution of micro-Jy radio sources}

We now analyze in more detail the evolution of the various classes and
compare it with previous results. 

\subsubsection{Star-forming galaxies}\label{sfg_evol}

Our results on the evolution of SFG in the radio band, which we fitted with
$P(z) \propto (1+z)^{k_L}$, $k_L = 2.89^{+0.10}_{-0.15}$ (or $k_L =
2.5^{+0.2}_{-0.3}$ from the $\langle V_{\rm e}/V_{\rm a} \rangle$ method)
in the range $0 \la z \le 2.3$, agree very well with previous
determinations. For example, \cite{hop04} obtained $k_L=2.7\pm0.6$ up to
$z=2$ (and constant thereafter), with a small (but not significant)
component of density evolution $\Phi(z) \propto (1+z)^{0.15\pm0.60}$. To
check if we could constrain a possible redshift peak in the evolution we
tried a simple model of the type $P(z) = (1+z)^{k+\beta z}$, first
suggested by \cite{wa08}, which allows for a maximum in the luminosity
evolution followed by a decline. We found no evidence for $\beta$ being
significantly different from 0, but we should stress that only two of our
SFG have $z > 2$.

As regards other bands, \cite{mag09} have modeled the evolution of
infrared luminous star-forming galaxies as a pure luminosity evolution
$P(z)\propto (1+z)^{3.6\pm0.4}$ up to $z \sim 1.3$, while \cite{mag11} have
derived $P(z) \propto (1+z)^{1.0\pm0.9}$ for $1.3 \la z \la 2.3$, which
suggests a slowing down of the evolution at $z \ga 1.3$ (in both cases the
evidence for density evolution is not significant).
If we split our sample we derive $k_L = 3.5^{+0.4}_{-0.7}$ ($\langle V_{\rm
  e}/V_{\rm a} \rangle = 0.63\pm0.04$, $P_{\rm ev} = 99.98\%$) for $z \le
1.3$ ($k_L = 3.1^{+0.8}_{-1.0}$ excluding the two large-scale structures)
and $k_L = 1.6^{+0.6}_{-0.7}$ ($\langle V_{\rm e}/V_{\rm a} \rangle =
0.66\pm0.09$, $P_{\rm ev} \sim 54\%$) for $1.3 \le z \le 2.3$, which is
consistent with the IR results.  In summary, although we find no
significant evidence for a slowing down of the evolution at higher
redshifts, such an occurrence cannot be ruled out by our data.

Our SFG have $\langle L_{\rm IR} \rangle \sim 5 \times 10^{11} L_{\odot}$,
which is typical for luminous infrared galaxies (LIRGs; $10^{11} \le L_{\rm
  IR} \le 10^{12} L_{\odot}$), and extend over $8 \times 10^{9} \la L_{\rm
  IR} \la 2 \times 10^{13} L_{\odot}$ \citep[where $L_{\rm IR}$ refers to
the $8 - 1,000~\mu$m range and we have used the mean value of the $L_{\rm
  IR}/P_{1.4GHz}$ ratio given in][to estimate it]{sar10}, thereby reaching
well into the ultra-luminous infrared galaxies (ULIRGs) regime ($L_{\rm
  IR} > 10^{12} L_{\odot}$).

\subsubsection{Radio-loud AGN}\label{sect:rlagn}

We have found a significant difference between the evolutionary properties
of low-power and high-power radio-loud AGN, with a dividing line between
the two at $P \sim 3 \times 10^{24}$ W Hz$^{-1}$. Namely, while low-power
sources do not evolve, high-power ones anti-evolve significantly ($P \sim
99.9\%$), which indicates that they were either less numerous or less
luminous in the past. This is exactly the opposite of what was found in samples
at higher flux densities ($\ga 1$ Jy), where high-power sources exhibit a
strong {\it positive} evolution, with low-power ones still not evolving 
\cite[e.g.,][]{up95,ja99}. This difference is likely due to two
factors: 1.
the range of redshifts sampled in the two cases are very different. For
example, while a $10^{25}$ W~Hz$^{-1}$ source in a sample defined by
$S_{1.4GHz,min} \ge 1$ Jy can be detected only up to $z \sim 0.065$, if
$S_{1.4GHz,min} \ge 100~\mu$Jy the same source can be seen up to $z \sim
4.1$ ($\alpha_{\rm r} = 0.7$). And indeed, for our high-power radio-loud
AGN $\langle z \rangle \sim 2$ (Tab. \ref{tabveva}), which is where these
sources are supposed to start their decline \citep[e.g.,][]{wad01}; 2. a
class of sources with $P \ga 3 \times 10^{24}$ W Hz$^{-1}$, moderate
luminosity evolution, and a cutoff redshift $z_{\rm max} = 5.5$, will reach
a limiting flux density $f_{\rm min} \approx 70~\mu$Jy \citep{pad11}.  When
the flux density limit $S_{min}$ of a survey is comparable to, or even
lower than, this value, the observer will start ``running out'' of sources
and a deficit at higher redshifts will be observed.  Stated differently,
since at a first order $V_{\rm e}/V_{\rm a} \sim (S/S_{min})^{-3/2}$, in
this case $S/S_{min}$ can often be $>1$, if not $\gg 1$, which translates
into small values of $V_{\rm e}/V_{\rm a}$. The key assumption here is
that there needs to be a redshift cutoff, as if sources were present at all
redshifts then $f_{\rm min}$ would be tending to zero and no such effect
would be present.

Note that the radio power, which separates the non-evolving from the
anti-evolving radio sources coincides with the minimum power of FR II
sources \citep[e.g.,][]{up95,gen10,pad11} and radio-quasars \citep{pad11}.
We then identify our low-power radio-loud AGN with low-luminosity radio
galaxies of the FR I type and the high-power ones with the FR II-like,
powerful radio sources, which dominate the bright ($\ga 1$ mJy) radio sky,
and of which we are witnessing the demise.

\cite{sad07} and \cite{do09} have studied the radio evolution of luminous
red galaxies up to $z = 0.7$ and $0.8$, finding evidence of weak but
significant positive evolution. Their samples have very little overlap with
ours, as they include radio sources with $P \ge 10^{24.2} - 10^{24.3}$ W
Hz$^{-1}$ and only seven ($\sim 8\%$) of our AGN in this power
range have $z < 0.8$ (see Fig. \ref{prz}). Furthermore, both studies only
deal with red galaxies, while we include all radio sources. Our LF in the
$0.4 - 0.8$ redshift range is in any case fully consistent with theirs,
although our uncertainties are obviously quite large, once we exclude four
sources belonging to the two large scale structures discussed in
Section~\ref{lss} and one optically compact source (which would have not
been included by either of the two studies). The small, positive evolution
they detect even at the lowest powers, that is between $10^{24.2}$ and
$10^{25.8}$ W Hz$^{-1}$ and $10^{24.3}$ and $10^{25}$ W Hz$^{-1}$
respectively, is consistent with these ranges encompassing the minimum
value for FR II radio galaxies, which are known to evolve positively at the
radio flux densities sampled by both studies \citep[e.g.,][]{gen10}.

\cite{tay09} have found that the number density of massive ($M_{\star} >
10^{11} M_{\odot}$) red galaxies declines with redshift as $\Phi(z) \propto
(1+z)^{-1.60\pm0.14(\pm0.21)}$ for $z \le 1.8$. This is tantalizingly
similar to the dependence we find for our high-power radio-loud AGN
$\Phi(z) \propto (1+z)^{-1.8\pm0.6}$ over a larger redshift range ($0.5 -
5.8$). If we restrict our sample to $z \le 1.8$ we still find evidence of
negative evolution but not significantly so due to the smaller sample size.

\subsubsection{Radio-quiet AGN}

We have estimated for the first time the evolution of radio-quiet AGN in
the radio band, modeling it as a pure luminosity evolution and obtaining
$P(z) \propto (1+z)^{k_L}$, $k_L = 2.5^{+0.4}_{-0.5}$, in the range $0.2 \la
z \la 3.9$. In the X-ray band the situation appears to be more complex,
with strong evolution up to $z \sim 1 - 2$ (depending on luminosity) and
then a slow down \citep[e.g.,][]{ha05}.
As done for SFG, we then tried a model of the type $P(z) = (1+z)^{k+\beta
  z}$ but our previous best fit ($k = 2.5$, $\beta = 0$) was still fully
consistent with the data, although peaks at $z \approx 3$, for example, are
also within the $1\sigma$ contours. Larger samples of radio-selected,
radio-quiet samples will put strong constraints on the evolution of these
sources, by-passing also the problems of obscuration, which plague the
optical and soft X-ray bands, although source identification will require
ancillary multi-wavelength data, as shown in Section~\ref{sample}.
Moreover, the surface density of our radio-quiet AGN, $\sim 520$
deg$^{-2}$, is already a factor $\sim 6$ larger than that of one of the
deepest optically-selected quasar samples \citep[$\sim 80$ deg$^{-2}$ to $g
  \sim 22$;][]{ric05} and only $\sim 1/4$ of that of unabsorbed AGN down to
faint Chandra fluxes \citep[$\sim 2,000$ deg$^{-2}$;][]{ha05}. (Note
however that we are sensitive to both broad- and narrow-lined AGN, while
both the optical and X-ray samples under consideration include only the
former type.)

Our radio number counts are consistent with those in the hard X-ray band,
as shown in Fig.  \ref{newcounts}, which shows the predictions of
\cite{wil08}, based on a conversion of the AGN X-ray LF to a radio LF, and
of Paper IV, obtained from the X-ray number counts by using a typical
radio-to-X-ray flux density ratio. This shows that the sources we are
selecting in the radio band are the same as the X-ray emitting radio-quiet
AGN.

%\begin{figure}%14
%%\includegraphics[scale=0.8]{Evolution_AGN_cropped.eps}
%\plotone{Evolution_AGN_new.eps}
%\caption{The evolution of sub-mJy AGN.}
%\label{summary_ev}
%\end{figure}

%Fig. \ref{summary_ev} summarizes the evolution of the various classes of AGN. 

%Sey Radio LF: resolution effects? A slight offset ($\sim 0.5$ in log) 
%would make our LF and the Sey one in good agreement ... 

\subsection{Comparison with the COSMOS and Deep SWIRE Field
  surveys}\label{comp_COSMOS}

\cite{smo09a} and \cite{smo09b} have studied the evolution of SFG and AGN
in the radio band using 1.4 GHz VLA observations of the COSMOS 2 deg$^2$
field, which have a limiting flux density $\sim 45~\mu$Jy in the central 1
deg$^2$. \cite{str10} have studied a 1.4 GHz selected sample of radio
sources in the Deep SWIRE Field (DSF), reaching a limiting flux density
$\sim 13.5~\mu$Jy at the center of a 0.36 deg$^2$ area. We can then compare
our results more directly with theirs, keeping in mind that both samples
reach only $z = 1.3$ and that quasars, which make up $\la 20\%$ of
the total, were not included in the AGN COSMOS sample.

\cite{smo09a} derived $k_L = 2.1\pm0.2$ or $k_L = 2.5\pm0.1$ for SFG
depending on the choice of the local LF, while we get $k_L =
3.5^{+0.4}_{-0.7}$ for $z \le 1.3$ ($k_L = 3.1^{+0.8}_{-1.0}$ excluding the
two large-scale structures). Their values are smaller than ours but not
significantly so, given our relatively large uncertainties and the presence
of large-scale structures in our field. Their evolution, unlike ours, is
however significantly weaker than found by \cite{mag09} in the IR band for $z \le 1.3$
($k_L = 3.6\pm0.4$: see Section~\ref{sfg_evol}). \cite{str10} have defined two
classes of ``star-forming'' (blue) and ``intermediate'' (green) galaxies,
for which they derive $k_L = 2.9\pm0.3$ and $k_L = 3.6\pm0.2$ respectively
(their non-parametric results). This evolution is stronger than obtained by
\cite{smo09a} but in good agreement with ours and also \cite{mag09}.

\cite{smo09b} found $k_L = 0.8\pm0.1$ or $k_D = 1.1\pm0.1$ for their AGN.
For $z \le 1.3$ we obtain $\langle V_{\rm e}/V_{\rm a} \rangle =
0.54\pm0.03$, indicative of slightly positive but not significant ($P_{\rm
  ev} = 77\%$) evolution, with $k_L = 1.5$ and $k_D = 1.0$ (and obviously
large error bars). If we exclude the two large-scale structures $\langle
V_{\rm e}/V_{\rm a} \rangle = 0.57\pm0.03$, indicative of positive ($P_{\rm
  ev} = 96.9\%$) evolution, with $k_L = 2.3\pm1.0$ and $k_D
=2.0\pm1.0$. These are larger than the COSMOS values but still consistent
with them given our relatively large error bars. Part of the difference
might also be explained by the exclusion of quasars from the COSMOS
sample, as these are expected to be mostly of the radio-quite type (Paper
IV) and therefore strongly evolving.  Indeed, when splitting the sample
into radio-quiet and radio-loud AGN one obtains $\langle V_{\rm e}/V_{\rm
  a} \rangle = 0.69\pm0.06$, $k_L = 3.9^{+0.7}_{-0.9}$ ($P_{\rm ev} \sim
96.8\%$) and $\langle V_{\rm e}/V_{\rm a} \rangle = 0.49\pm0.04$ ($P_{\rm
  ev} \sim 85\%$) respectively ($\langle V_{\rm e}/V_{\rm a} \rangle =
0.52\pm0.04$ excluding the two large-scale structures). The apparently
positive AGN evolution appears then to be driven by the radio-quiet
sources, while the radio-loud ones do not evolve in the redshift range
sampled by \cite{smo09b}. This has important implications for their
results, as they do not distinguish between the two classes of AGN and use
the positive evolution they found to estimate, for example, the evolution
of the comoving radio luminosity density and mechanical energy output of
low-radio-power AGN. \cite{str10} derived $k_L = 2.5\pm0.3$ for their
``quiescent'' (red) galaxies, which they identify as AGN, which is in good
agreement with our result but significantly larger than the COSMOS
value. Note that $\sim 40\%$ of the AGN in \cite{str10} are within
$1\sigma$ of the infrared-radio correlation typical of SFG, which shows
that, like ours, about half of their AGN {\it cannot} be of the radio-loud
type \citep[see also][]{pra10}.

\subsection{The origin of radio emission in radio-quiet AGN}\label{rq_emission}

The mechanism responsible for the radio emission in radio-quiet AGN has
been a matter of debate since the discovery of quasars. Alternatives have
included a scaled down version of the radio-loud AGN mechanism
\citep[e.g.,][]{mi93}, star formation \citep[e.g.,][]{so91}, a magnetically
heated corona \citep{la08}, and disk winds \citep{blu07} \citep[but
see][]{ste11}.

Our results suggest very close ties between star formation and radio
emission in radio-quiet AGN, since their evolution is indistinguishable
from that of SFG (Section~\ref{evolution}) and their LF appears to be an
extension of the SFG LF (Section~\ref{sect:LF}). Furthermore, radio
emission in the two classes of AGN is bound to have a different origin. If
radio-quiet AGN were simply ``mini radio-loud'' AGN, in fact, they would
have to share the evolutionary properties of the latter and their LF should
also be on the extrapolation of the radio-loud one at low powers. None of
these two facts is borne out by our data (see, e.g., Fig.
\ref{sfg_rq_rl}).

This concurs with the results of various papers over the past 20 years
\citep[e.g.,][]{so91,sar10}, which have shown that radio-quiet AGN and
star-forming galaxies have very similar FIR-to-radio flux density
ratios. Note, however, that \cite{sa89} have discarded this as entirely
coincidental, and similarly \cite{ku98} have suggested that dust heating by
the quasar and AGN-related radio emission could also conspire to make this
happen. Moreover, the detection of compact, high brightness temperature
cores in several radio-quiet AGN \citep[e.g.,][]{ul05}, which resemble
those observed in radio-loud AGN, would also argue against our results,
although in some Seyfert galaxies these cores are surrounded by diffuse
radio emission connected to star-forming regions \citep{or10}.

This suggests that AGN and star-formation related processes coexist in
radio-quiet AGN. Indeed, the fraction of flux density contained in the
compact, central component is, on average, $\sim 70\%$ for low-redshift
radio-quiet AGN \citep[e.g.,][]{kel89,ku98}, which leaves some room for
extended emission. If the AGN-related component is non-evolving, as appears
to be the case for low-power radio-loud AGN, while the
star-formation-related one follows the evolution of SFG, one could
understand the difference between our radio-quiet AGN, which have $\langle
z \rangle \sim 1.7$, and those imaged with the VLBI, which are mostly
local. In fact, since this redshift difference implies an increase by a
factor $\approx 10 - 20$ in the radio power related to star formation, a
minor ($\sim 1/3$) extended component at $z \sim 0$ would then
become dominant ($\ga 3$) at higher redshifts. Furthermore, it should be pointed
out that no complete sample of radio-selected, radio-quiet AGN has ever
been observed at the VLBI resolution. The choice of objects that could be
detected might have then led to a selection effect favoring relatively
strong targets, more similar to radio-loud sources.

\cite{ric07} have studied a sample of 92 radio sources in the Hubble Deep
Field North brighter than $40~\mu$Jy and well resolved by MERLIN and the
VLA.  By classifying more than 70\% of them as starbursts or AGN using
radio morphologies, spectral indices, optical appearance, and rest-frame
MIR emission, they found that the X-ray luminosity indicates the presence
of an AGN in at least half of the 45 radio starbursts with X-ray
counterparts. Moreover, almost all extended radio starbursts at $z > 1.3$
host X-ray selected obscured AGN and their radio and X-ray powers are
uncorrelated, which points to different emission mechanisms being at play
in the two bands. These results, which associate high-redshift radio
starbursts with AGN, are fully consistent with our suggestion of a very
close relationship between star formation and radio emission in our
relatively high-redshift radio-quiet AGN.

If radio emission in radio-quiet AGN is mostly related to star formation
processes, we can use the mean value of the $L_{\rm IR}/P_{1.4GHz}$ ratio
given in \cite{sar10} to estimate their IR powers.  Our radio-quiet AGN
have $\langle L_{\rm IR} \rangle \sim 4 \times 10^{12} L_{\odot}$, that is
well in the ULIRGs regime, and reach $2 \times 10^{14} L_{\odot}$. The
star-formation rates (SFRs) implied by their radio powers assuming that all
radio emission is star-formation related and using the relationship derived
by \cite{be03}, are, on average, $\sim 500~M_{\odot}$ yr$^{-1}$, extending
over the $\sim 10 - 20,000~M_{\odot}$ yr$^{-1}$ range. The mean value is
typical of ULIRGs, while the upper end is in the hyperluminous infrared
galaxies regime ($L_{\rm IR} > 10^{13} L_{\odot}$), which can reach SFRs
$>10,000~M_{\odot}$ yr$^{-1}$ \citep[e.g.,][]{ro10}.

The association of ULIRGs with radio-quiet AGN is certainly not
new. \cite{sa88} proposed an evolutionary connection between ULIRGs and
quasars, based on the fact that all of the twelve ULIRGs in their sample
displayed AGN spectra in the optical band ($\sim 2/3$ of their quasars are
radio-quiet). Moreover, the AGN detection rate amongst local ULIRGs is
$\sim 70\%$ \citep{na10} and radio-detected ULIRGs are known to be 
rare locally but rapidly evolving with redshift \citep[e.g.,][]{mau07}.

\subsection{Astrophysics of micro-Jy sources}\label{astro}

\cite{fan74} recognized that radio galaxies separate into two distinct
luminosity classes, each with its own characteristic radio morphology.
High-luminosity FR IIs have radio lobes with prominent hot spots and bright
outer edges, while in low-luminosity FR Is radio emission is more
diffuse. The luminosity distinction is fairly sharp at 178 MHz, with FR Is
and FR IIs lying below and above, respectively, the fiducial luminosity
$P_{\rm 178 MHz} \approx 10^{26}/(H_0/70)^2$ W Hz$^{-1}$. This translates
at higher frequencies to $P_{\rm 1.4 GHz} \approx 3 \times
10^{25}/(H_0/70)^2$ W Hz$^{-1}$ (assuming $\alpha_{\rm r} = 0.7$), with
some dependency also on optical luminosity \citep{ow91} and therefore a
rather large overlap. An independent separation on the basis of nuclear
activity into high-excitation (HERGs) and low-excitation (LERGs)
radio-galaxies has been proposed more recently \citep{la94}. It turns out
that almost all FR Is are LERGs and most FR IIs are HERGs, although there
is a population of FR II LERGs as well. Observational evidence indicates
that the two types of radio galaxies have intrinsically different central
engines. Namely in LERGs the accretion disk, if at all present, is thought
to be much less efficient than in HERGs \citep[e.g.,][]{chi99,do04,ev06}. 
This points to a large difference in accretion rates between the two
classes.

\cite{cro06} have associated high-accretion sources, and therefore also
HERGs, with their so-called ``quasar-mode'', which they interpret as
merger-driven, efficient accretion of cold disk gas, present also in
radio-quiet AGN. Low-accretion objects, i.e., LERGs, on the other hand,
have been connected with the less efficient accretion of warm gas, the
so-called ``radio-mode''. The mean black hole accretion rate for the
``radio-mode'' is predicted to be approximately constant up to $z \approx
2$, based on a suite of semi-analytic models implemented on the output of
the Millennium Run \citep[Fig. 3 of][]{cro06}. ``Quasar-mode'' accretion,
on the other hand, is envisioned to be most efficient at $z \sim 2 - 4$,
dropping by a factor of 5 by $z \sim 0$. This is similar in form but
somewhat weaker than the observed cosmological luminosity evolution of
bright quasars in the radio, optical, and X-ray bands \citep[e.g.,][and
  references therein]{wa05}.

%optical ev.: Croom et al.: 4.4 between 0 and 2 (Tab. 5, first line)
%radio ev.: Padovani et al.: 7.3 between 0 and ~ 2 (Fig. 10)
 %" '' : Wall et al. (2005): ~ 10 between 0 and ~ 2 (Fig. 3) 
% X-ray: Hasinger et al.: ~ 14 between 0 and 1.7 (Tab. 5) 

As we identify our low-power radio-loud AGN with FR Is/LERGs these would 
then be inefficient accretors in a ``radio-mode'', which would explain their 
lack of evolution, assuming that it is driven only by the accretion rate. 
We instead identify as sources in a ``quasar-mode'' the high-power 
radio-loud AGN (which should be mostly HERGs) and the radio-quiet ones. 
Their different evolutions in the radio band are explained by distinct 
emission mechanisms and by the fact that the high radio powers of the 
former mean that we are seeing the effects of a high redshift cutoff.

\section{Summary and Conclusions}

We have used a deep, complete radio sample of 193 objects down to a 1.4 GHz
flux density of $43~\mu$Jy selected in the Chandra Deep Field South area to
sharpen our understanding of the nature of sub-millijansky sources and to study for
the first time their evolution and luminosity functions up to $z \sim
5$. Our unique set of ancillary data, which includes far-IR, near-IR, and
optical observations, redshift information, and X-ray detections or upper
limits for a large fraction of our sources, has allowed us to develop an
unprecedented classification scheme to categorize in a robust way faint
radio sources in star-forming galaxies, radio-quiet, and radio-loud AGN.
Our main results can be summarized as follows:

\begin{enumerate}
\item Star-forming galaxies and AGN make up an equal part of the sub-millijansky
  sky down to $43~\mu$Jy, with the former becoming the dominant population
  only below $\approx 0.1$ mJy. Radio-quiet AGN are confirmed to be an
  important class of sub-millijansky sources, accounting for $\sim 30\%$ of the
  sample and $\sim 60\%$ of all AGN, and outnumbering radio-loud AGN at
  $\la 0.1$ mJy.
\item The radio power of star-forming galaxies evolves as $(1+z)^{2.5 -
  2.9}$ up to $z \le 2.3$, their maximum redshift in our sample, in
  agreement with previous determinations in the radio and IR
  bands. Although evidence of a slowing down of the evolution at $z \ga
  1.3$ is not significant it cannot also be ruled out. The radio luminosity
  function of SFG can be parametrized as broken power-law $\Phi(P) \propto
  P^{-1.3}$ and $\Phi(P) \propto P^{-3.15}$ at the faint and bright end
  respectively, with a break at $P \sim 7 \times 10^{21}$ W~Hz$^{-1}$,
  which is also consistent with previous derivations.
\item AGN as a whole do not appear to evolve. However, once they are split
  into radio-quiet (energy budget dominated by thermal emission) and
  radio-loud (dominated by non-thermal, jet emission) sources the situation
  is very different, with the radio-quiet population evolving very
  significantly and similarly to star-forming galaxies and the radio-loud
  population displaying negative density evolution $\Phi(z) \propto
  (1+z)^{-1.8\pm0.4}$. The luminosity function of radio-loud AGN, $\Phi(P)
  \propto P^{-1.5}$, is also much flatter than that of radio-quiet AGN,
  $\Phi(P) \propto P^{-2.6}$, derived here for the first time, which seems
  to be an extension of that of star-forming galaxies at the high power
  end.
\item There is a significant difference between the evolutionary properties
  of low-luminosity radio galaxies and radio powerful ($P \ga 3 \times
  10^{24}$ W Hz$^{-1}$) AGN, as while the former do not evolve, the latter
  evolve negatively. This is exactly the opposite of what found in samples
  at higher flux densities ($\ga 1$ Jy), where high-power sources exhibit a
  strong positive evolution. We interpret this difference as due to the
  fact that we are sampling the high-power radio-loud population up to much
  larger redshifts ($z \ga 5$) and as a result of a redshift cutoff.
\item Our results suggest a very close relationship between star formation
  and radio emission in radio-quiet AGN, since their evolution and
  luminosity function are respectively indistinguishable from, and an
  extension of, that of star-forming galaxies. This is supported by the
  fact that radio-quiet AGN and star-forming galaxies appear both to follow
  the ``IR-radio relation'' but is in contradiction with the detection of
  compact, high brightness temperature cores in several (mostly local) radio-quiet AGN,
  similar to those observed in radio-loud ones. The co-existence of two
  components, one non-evolving and AGN-related, and one evolving and
  star-formation-related, and selection effects in the choice of
  radio-quiet VLBI targets, can reconcile these apparently discrepant
  observational data.
\item The surface density of radio-selected, radio-quiet AGN, $\sim 520$
  deg$^{-2}$, is already about 6 times larger than that of one of the
  deepest optically-selected quasar sample and only $\sim 1/4$ that of
  unabsorbed X-ray selected AGN. This means that sub-millijansky radio 
  surveys, given the appropriate ancillary multi-wavelength data, have
  the potential of detecting large numbers of radio-quiet AGN by-passing
  the problems of obscuration, which plague the optical and soft X-ray
  bands. The radio number counts of radio-quiet AGN are consistent with
  those in the hard X-ray band, which shows that the sources we are
  selecting are the same as the X-ray emitting radio-quiet AGN.
\item Sub-millijansky radio surveys wanting to study the evolution of faint, {\it
  radio-loud} AGN, need to consider that a large fraction ($\sim 60\%$ down
  to $\sim 50~\mu$Jy) of the radio-selected AGN are actually of the {\it
    radio-quiet} type and therefore need to be treated separately. This has
  not been done so far by other studies and can have a large impact on the
  study of faint ``radio-mode'' inefficient accretors.

\end{enumerate}

We plan to expand on this work by using our deeper radio observations 
\citep[][and in preparation]{mil08} and the recently released 4 Msec 
Chandra data \citep{xue11}. This will provide us with a catalogue of $\sim 900$
radio sources, with which we will be able to address the issues 
discussed in this paper with larger statistics. 

\begin{acknowledgements}
  NM acknowledges the support of a Chandra Award AR8-9016X. PT acknowledges
  support from the ASI--INAF I/009/10/0 grant. We thank Evanthia
  Hatziminaoglou for useful discussions and Mark Dickinson and the FIDEL
  Team for providing the $70~\mu$m FIDEL image and catalogue. Extensive use
  was made of the TOPCAT software package \citep{tay05}. We acknowledge the
  ESO/GOODS project for the ISAAC and FORS2 data obtained using the Very
  Large Telescope at the ESO Paranal Observatory under Program ID(s):
  LP168.A-0485, 170.A-0788, 074.A-0709, 275.A-5060, and 081.A-0525.  The
  VLA is a facility of the National Radio Astronomy Observatory which is
  operated by Associated Universities, Inc., under a cooperative agreement
  with the National Science Foundation. This research has made use of the
  NASA/IPAC Extragalactic Database (NED) which is operated by the Jet
  Propulsion Laboratory, California Institute of Technology, under contract
  with the National Aeronautics and Space Administration and NASA's
  Astrophysics Data System (ADS) Bibliographic Services.

{\it Facilities:} \facility{VLA}, \facility{ESO:3.6m (EFOSC2)},
\facility{VLT:Kueyen (FORS1)}, \facility{VLT:Antu (FORS2)},
\facility{VLT:Antu (ISAAC)}, \facility{Max Plank:2.2m (WFI)},
\facility{CXO}, \facility{XMM}, \facility{Spitzer (IRAC)},
\facility{Spitzer (MIPS)}

%ARE THERE MISSING FACILITIES? 

\end{acknowledgements}

\end{document}